\documentclass[prb,twocolumn,superscriptaddress,preprintnumbers]{revtex4}
\pdfoutput=1
\usepackage{amssymb}
\usepackage{amsmath}
\usepackage{graphicx}
\usepackage{dcolumn}
\usepackage{bm}
\usepackage[bookmarks = true, pdfpagemode = None, pdfstartview = FitH,colorlinks = true,citecolor = black, linkcolor = black, urlcolor = black]{hyperref}
\usepackage{url}

\newcommand{\bra}[1]{\left\langle{#1}\right\vert}
\newcommand{\ket}[1]{\left\vert{#1}\right\rangle}
\newcommand{\iqp}{\left|I_q^p\right|}
\newcommand{\Phiqx}{\Phi^x_q}
\newcommand{\Phicjjx}{\Phi^x_{\text{cjj}}}
\newcommand{\Phiccjjx}{\Phi^x_{\text{ccjj}}}
\newcommand{\Philx}{\Phi^x_{L}}
\newcommand{\Phirx}{\Phi^x_{R}}
\newcommand{\Phicox}{\Phi^x_{\text{co}}}
\newcommand{\Phiqfpx}{\Phi^x_{\text{qfp}}}
\newcommand{\Philatchx}{\Phi^x_{\text{latch}}}

\begin{document}

\title{Experimental Demonstration of a Robust and Scalable Flux Qubit}
\author{R. Harris}
\author{J. Johansson}
\author{A.J. Berkley}
\author{M.W. Johnson}
\author{T. Lanting}
\affiliation{D-Wave Systems, 4401 Still Creek Drive, Burnaby BC, Canada, V5C 6G9}
\author{Siyuan Han}
\affiliation{Department of Physics and Astronomy, University of Kansas, Lawrence KS, USA, 66045}
\author{P. Bunyk}
\author{E. Ladizinsky}
\author{T. Oh}
\author{I. Perminov}
\author{E. Tolkacheva}
\author{S. Uchaikin}
\author{E. Chapple}
\author{C. Enderud}
\author{C. Rich}
\author{M. Thom}
\author{J. Wang}
\author{B. Wilson}
\author{G. Rose}
\affiliation{D-Wave Systems, 4401 Still Creek Drive, Burnaby BC, Canada, V5C 6G9}
\date{\today}

\begin{abstract}
A novel rf-SQUID flux qubit that is robust against fabrication variations in Josephson junction critical currents and device inductance has been implemented.  Measurements of the persistent current and of the tunneling energy between the two lowest lying states, both in the coherent and incoherent regime, are presented.  These experimental results are shown to be in agreement with predictions of a quantum mechanical Hamiltonian whose parameters were independently calibrated, thus justifying the identification of this device as a flux qubit.  In addition, measurements of the flux and critical current noise spectral densities are presented that indicate that these devices with Nb wiring are comparable to the best Al wiring rf-SQUIDs reported in the literature thusfar, with a $1/f$ flux noise spectral density at $1\,$Hz of $1.3^{+0.7}_{-0.5}\,\mu\Phi_0/\sqrt{\text{Hz}}$.  An explicit formula for converting the observed flux noise spectral density into a free induction decay time for a flux qubit biased to its optimal point and operated in the energy eigenbasis is presented.
\end{abstract}

\maketitle

\section{Motivation}

Experimental efforts to develop useful solid state quantum information processors have encountered a host of practical problems that have substantially limited progress.  While the desire to reduce noise in solid state qubits appears to be the key factor that drives much of the recent work in this field, it must be acknowledged that there are formidable challenges related to architecture, circuit density, fabrication variation, calibration and control that also deserve attention.  For example, a qubit that is inherently exponentially sensitive to fabrication variations with no recourse for in-situ correction holds little promise in any large scale architecture, even with the best of modern fabrication facilities.  Thus, a qubit designed in the absence of information concerning its ultimate use in a larger scale system may prove to be of little utility in the future.  In what follows, we present an experimental demonstration of a novel superconducting flux qubit \cite{fluxqubit} that has been specifically designed to address several issues that pertain to the implementation of a large scale quantum information processor.  While noise is not the central focus of this article, we nonetheless present experimental evidence that, despite its physical size and relative complexity, the observed flux noise in this flux qubit is comparable to the quietest such devices reported upon in the literature to date. 

It has been well established that rf-SQUIDs can be used as qubits given an appropriate choice of device parameters.  Such devices can be operated as a flux biased phase qubit using two intrawell energy levels \cite{FluxBiasedPhaseQubit} or as a flux qubit using any pair of interwell levels \cite{fluxqubit}.  This article will focus upon an experimental demonstration of a novel rf-SQUID flux qubit that can be tuned in-situ using solely {\it static} flux biases to compensate for fabrication variations in device parameters, both within single qubits and between multiple qubits.  It is stressed that this latter issue is of critical importance in the development of useful large scale quantum information processors that could foreseeably involve thousands of qubits \cite{DiVincenzo}.  Note that in this regard, the ion trap approach to building a quantum information processor has a considerable advantage in that the qubits are intrinsically identical, albeit the challenge is then to characterize and control the trapping potential with high fidelity \cite{Wineland}.  While our research group's express interest is in the development of a large scale superconducting adiabatic quantum optimization [AQO] processor \cite{AQC,Santoro}, it should be noted that many of the practical problems confronted herein are also of concern  to those interested in implementing gate model quantum computation [GMQC] processors \cite{GMQC} using superconducting technologies.

This article is organized as follows: In Section II, a theoretical argument is presented to justify the rf-SQUID design that has been implemented.  It is shown that this design is robust against fabrication variations in Josephson junction critical current.  Second, it is argued why it is necessary to include a tunable inductance in the flux qubit to account for differences in inductance between qubits in a multi-qubit architecture and to compensate for changes in qubit inductance during operation.  Thereafter, the focus of the article shifts towards an experimental demonstration of the rf-SQUID flux qubit.  The architecture of the experimental device and its operation are discussed in Section III and then a series of experiments to characterize the rf-SQUID and to highlight its control are presented in Section IV.  Section V contains measurements of properties that indicate that this more complex rf-SQUID is indeed a flux qubit.  Flux and critical current noise measurements and a formula for converting the measured flux noise spectral density into a free induction (Ramsey) decay time are presented in Section VI.  A summary of key conclusions is provided in Section VII.  Detailed calculations of rf-SQUID Hamiltonians have been placed in the appendices.

\section{rf-SQUID Flux Qubit Design}

The behavior of most superconducting devices is governed by three types of macroscopic parameters: the critical currents of any Josephson junctions, the net capacitance across the junctions and the inductance of the superconducting wiring.  The Hamiltonian for many of these devices can generically be written as
\begin{equation}
\label{eqn:Hphase}
{\cal H}=\sum_i\left[\frac{Q_i^2}{2C_i}-E_{Ji}\cos(\varphi_i)\right]+\sum_{n}U_n\frac{\left(\varphi_n-\varphi_n^x\right)^2}{2} \; ,
\end{equation}

\noindent where $C_i$, $E_{Ji}=I_i\Phi_0/2\pi$ and $I_i$ denote the capacitance, Josephson energy and critical current of Josephson junction $i$, respectively.  The terms in the first sum are readily recognized as being the Hamiltonians of the individual junctions for which the quantum mechanical phase across the junction $\varphi_i$ and the charge collected on the junction $Q_i$ obey the commutation relation $[\Phi_0\varphi_i/2\pi,Q_j]=i\hbar\delta_{ij}$.  The index $n$ in the second summation is over closed inductive loops.  External fluxes threading each closed loop, $\Phi_n^x$, have been represented as phases $\varphi_n^x\equiv 2\pi\Phi_n^x/\Phi_0$.  The quantum mechanical phase drop experienced by the superconducting condensate circulating around any closed loop is denoted as $\varphi_n$.  The overall potential energy scale factor for each closed loop is given by $U_n\equiv(\Phi_0/2\pi)^2/L_n$.  Here, $L_n$ can be either a geometric inductance from wiring or Josephson inductance from large junctions \cite{vanDuzer}.  Hamiltonian (\ref{eqn:Hphase}) will be used as the progenitor for all device Hamiltonians that follow.

\subsection{Compound-Compound Josephson Junction Structure}

\begin{figure}
\includegraphics[width=3.25in]{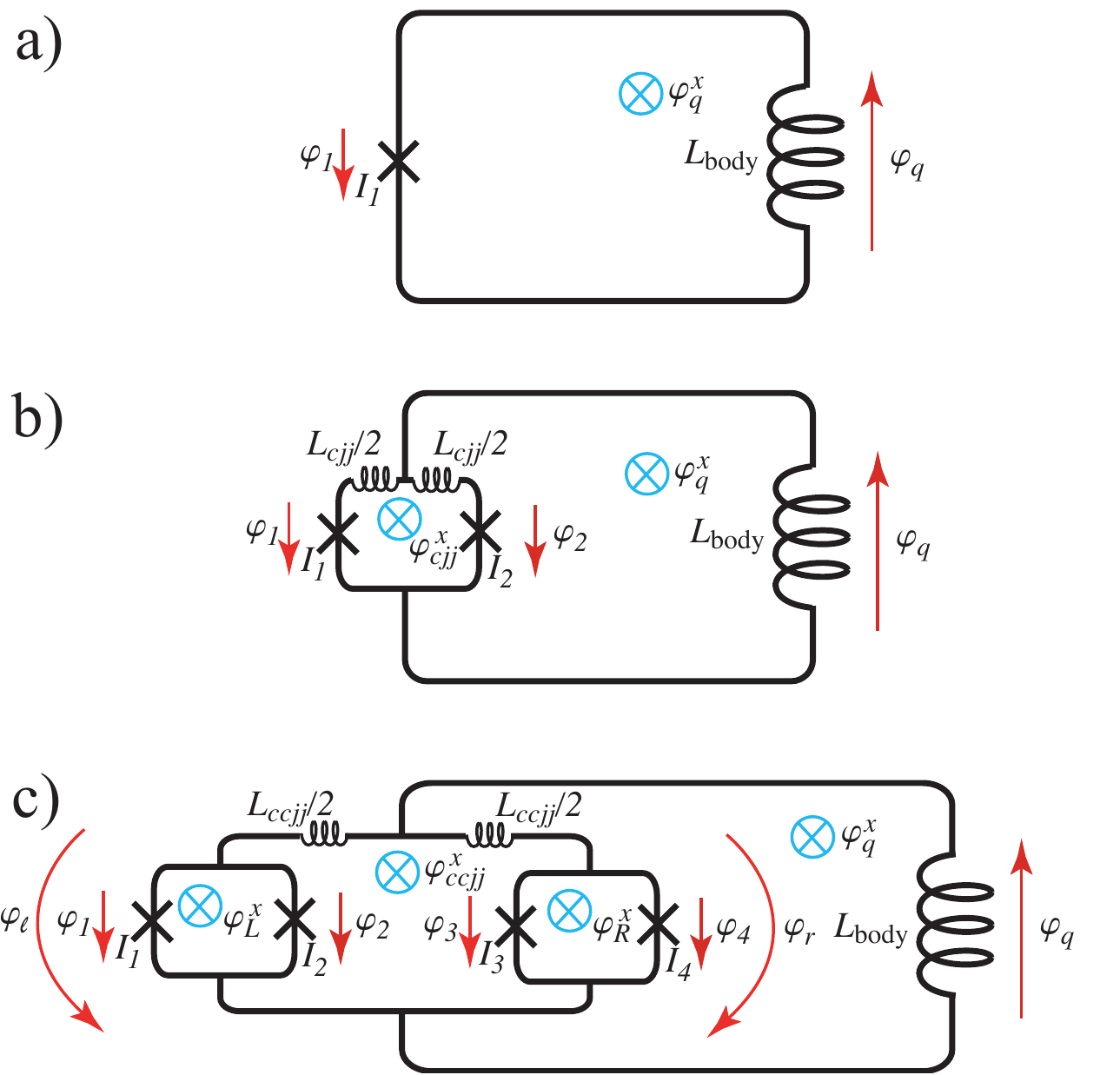}
\caption{\label{fig:rfssummary} (color online) a) A single junction rf-SQUID qubit.  b) Compound Josephson Junction (CJJ) rf-SQUID qubit.  c)  Compound-Compound Josephson Junction (CCJJ) rf-SQUID qubit.  Junction critical currents $I_i$ and junction phases $\varphi_i$ ($1\leq i \leq 4$) as noted.  Net device phases are denoted as $\varphi_{\alpha}$, where $\alpha\in\left(\ell ,r,q\right)$.  External fluxes, $\Phi_n^x$, are represented as phases $\varphi_{n}^x\equiv2\pi\Phi_n^x/\Phi_0$, where $n\in\left(L,R,\text{cjj},\text{ccjj},q\right)$.  Inductance of the rf-SQUID body, CJJ loop and CCJJ loop are denoted as $L_{\text{body}}$, $L_{\text{cjj}}$ and $L_{\text{ccjj}}$, respectively.}
\end{figure}

A sequence of rf-SQUID architectures are depicted in Fig.~\ref{fig:rfssummary}.  The most primitive version of such a device is depicted in Fig.~\ref{fig:rfssummary}a, and more complex variants in Figs.\ref{fig:rfssummary} b and \ref{fig:rfssummary}c.  For the single junction rf-SQUID (Fig.~\ref{fig:rfssummary}a), the phase across the junction can be equated to the phase drop across the body of the rf-SQUID: $\varphi_1=\varphi_q$.  The Hamiltonian for this device can then be written as
\begin{subequations}
\begin{equation}
\label{eqn:1JHeff}
{\cal H}=\frac{Q_q^2}{2C_q}+V(\varphi_q) \; ;
\end{equation}
\vspace{-0.12in}
\begin{equation}
\label{eqn:1JV}
V(\varphi_q)=U_q\Big\{\frac{\left(\varphi_q-\varphi_q^x\right)^2}{2}-\beta\cos\left(\varphi_q\right)\Big\} \; ;
\end{equation}
\vspace{-0.12in}
\begin{equation}
\label{eqn:1Jbeta}
\beta=\frac{2\pi L_q I_q^c}{\Phi_0} \; ,
\end{equation}
\end{subequations}

\noindent with the qubit inductance $L_q\equiv L_{\text{body}}$, qubit capacitance $C_q\equiv C_1$ and qubit critical current $I_q^c\equiv I_1$ in this particular case.  If this device has been designed such that $\beta>1$ and is flux biased such that $\varphi_q^x\approx\pi$, then the potential energy $V(\varphi_q)$ will be bistable.  With increasing $\beta$ an appreciable potential energy barrier forms between the two local minima of $V(\varphi_q)$, through which the two lowest lying states of the rf-SQUID may couple via quantum tunneling.  It is these two lowest lying states, which are separated from all other rf-SQUID states by an energy of order of the rf-SQUID plasma energy $\hbar\omega_p\equiv\hbar/\sqrt{L_qC_1}$, that form the basis of a qubit.  One can write an effective low energy version of Hamiltonian (\ref{eqn:1JHeff}) as \cite{Leggett}
\begin{equation}
\label{eqn:Hqubit}
{\cal H}_{q}=-{\frac{1}{2}}\left[\epsilon\sigma_z+\Delta_q\sigma_x\right] \;\; ,
\end{equation}

\noindent where $\epsilon=2\iqp\left(\Phi_q^x-\Phi_0/2\right)$,  $\iqp$ is the magnitude of the persistent current that flows about the inductive $q$ loop when the device is biased hard [$\epsilon\gg\Delta_q$] to one side and $\Delta_q$ represents the tunneling energy between the otherwise degenerate counter-circulating persistent current states at $\Phiqx=\Phi_0/2$.  

\begin{figure}
\includegraphics[width=3.25in]{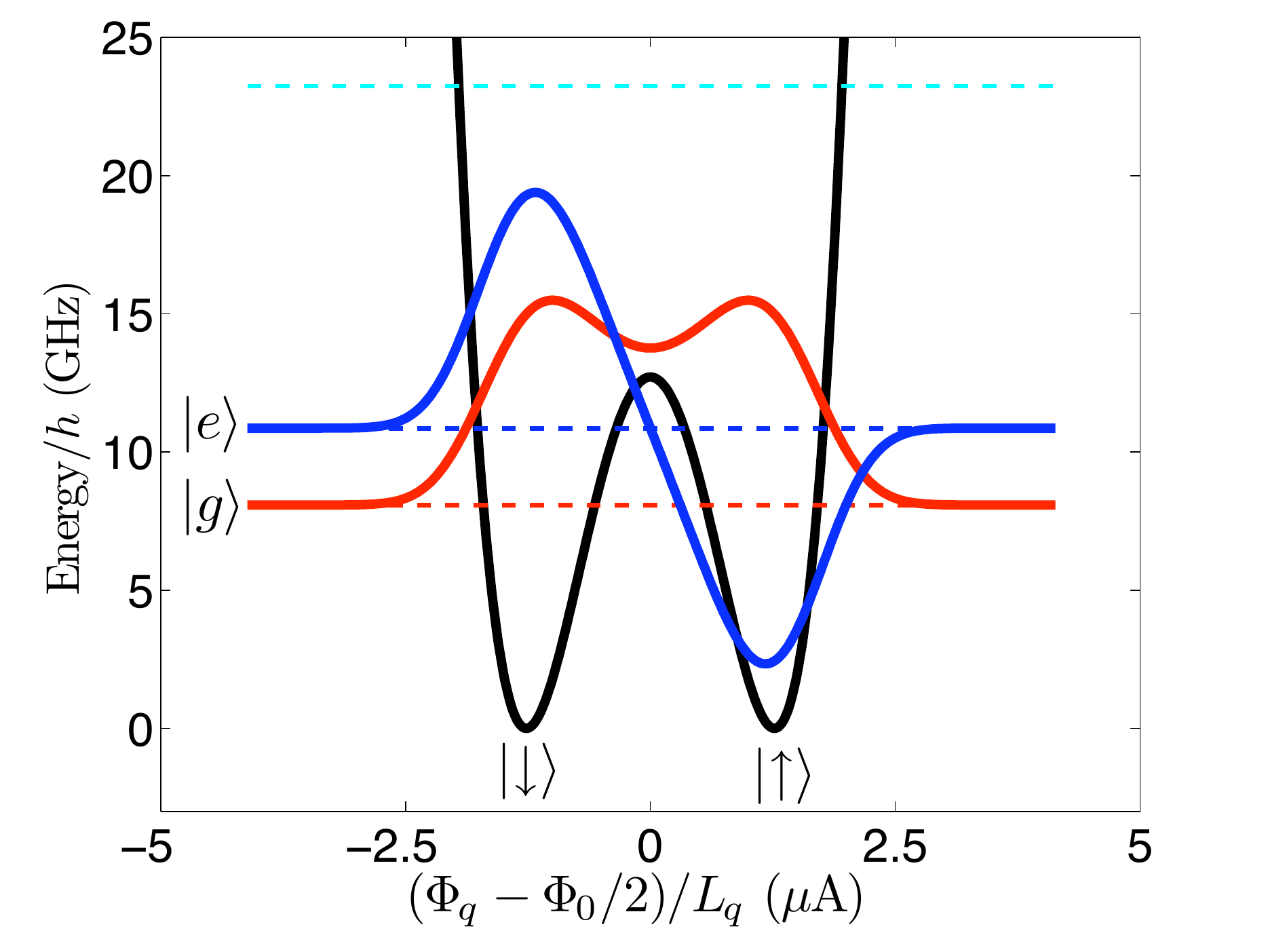}
\caption{\label{fig:qubitcomparison}  (color online) Depiction of the two lowest lying states of an rf-SQUID at degeneracy ($\epsilon=0$) with nomenclature for the energy basis ($\ket{g}$,$\ket{e}$) and flux basis ($\ket{\downarrow}$,$\ket{\uparrow}$) as indicated.}
\end{figure}

A depiction of the one-dimensional potential energy and the two lowest energy states of an rf-SQUID at degeneracy ($\Phi_q^x=\Phi_0/2$) for nominal device parameters is shown in Fig.~\ref{fig:qubitcomparison}.  In this diagram, the ground and first excited state are denoted by $\ket{g}$ and $\ket{e}$, respectively.  These two energy levels constitute the energy eigenbasis of a flux qubit.  An alternate representation of these states, which is frequently referred to as either the flux or persistent current basis, can be formed by taking the symmetric and antisymmetric combinations of the energy eigenstates: $\ket{\downarrow}=\left(\ket{g}+\ket{e}\right)/\sqrt{2}$ and $\ket{\uparrow}=\left(\ket{g}-\ket{e}\right)/\sqrt{2}$, which yield two roughly gaussian shaped wavefunctions that are centered about each of the wells shown in Fig.~\ref{fig:qubitcomparison}.  The magnitude of the persistent current used in Eq.~(\ref{eqn:Hqubit}) is then defined by $\iqp\equiv\left|\bra{\uparrow}\left(\Phi_q-\Phi_0/2\right)/L_q\ket{\uparrow}\right|$.  The tunneling energy is given by $\Delta_q=\bra{e}{\cal H}_q\ket{e}-\bra{g}{\cal H}_q\ket{g}$. 

The aforementioned dual representation of the states of a flux qubit allows two distinct modes of operation of the flux qubit as a binary logical element with a logical basis defined by the states $\ket{0}$ and $\ket{1}$.  In the first mode, the logical basis is mapped onto the energy eigenbasis: $\ket{0}\rightarrow\ket{g}$ and $\ket{1}\rightarrow\ket{e}$.  This mode is useful for optimizing the coherence times of flux qubits as the dispersion of Hamiltonian (\ref{eqn:Hqubit}) is flat as a function of $\Phi_q^x$ to first order for $\epsilon\approx 0$, thus providing some protection from the effects of low frequency flux noise \cite{optimalpoint}.  However, this is not a convenient mode of operation for implementing interactions between flux qubits \cite{parametriccoupling1,parametriccoupling2}.  In the second mode, the logical basis is mapped onto the persistent current basis: $\ket{0}\rightarrow\ket{\downarrow}$ and $\ket{1}\rightarrow\ket{\uparrow}$.  This mode of operation facilitates the implementation of inter-qubit interactions via inductive couplings, but does so at the expense of coherence times.  GMQC schemes exist that attempt to leverage the benefits of both of the above modes of operation \cite{IBM,Oliver,NiftyItalianPaper}.  On the other hand, those interested in implementing AQO strictly use the second mode of operation cited above.  This, very naturally, leads to some interesting properties:  First and foremost, in the coherent regime at $\epsilon=0$, the groundstate maps onto $\ket{g}=\left(\ket{0}+\ket{1}\right)/\sqrt{2}$, which implies that it is a superposition state with a fixed phase between components in the logical basis.  Second, the logical basis is not coincident with the energy eigenbasis, except in the extreme limit $\epsilon/\Delta_q\gg 1$.  As such, the qubit should not be viewed as an otherwise free spin-1/2 in a magnetic field, rather it maps onto an Ising spin subjected to a magnetic field with both a longitudinal ($B_z\rightarrow\epsilon$) and a transverse ($B_x\rightarrow\Delta_q$) component \cite{Ising}.  In this case, it is the competition between $\epsilon$ and $\Delta_q$ which dictates the relative amplitudes of $\ket{\downarrow}$ and $\ket{\uparrow}$ in the groundstate wavefunction $\ket{g}$, thereby enabling logical operations that make {\it no} explicit use of the excited state $\ket{e}$.  This latter mode of operation of the flux qubit has connections to the fields of quantum magnetism \cite{Anderson} and optimization theory \cite{Kirkpatrick}.  Interestingly, systems of coupled flux qubits that are operated in this mode bear considerable resemblance to Feynman's original vision of how to build a quantum computer \cite{Feynman}.

While much seminal work has been done on single junction and the related 3-Josephson junction rf-SQUID flux qubit\cite{3JJfluxqubits,MooijMore3JJFluxQubits,MooijSuperposition,MooijCoherentDynamics,MooijCoupledSpectroscopy,OliverMachZehnder,OliverLandauZener,OliverAmplitudeSpectroscopy,ClarkeQubits,IPHT4Q,1OverFFluxQubit1,1OverFFluxQubit2}, it has been recognized that such devices would be impractical in a large scale quantum information processor as their properties are exceptionally sensitive to fabrication variations.  In particular, in the regime $E_{J1}\gg\hbar\omega_p$, $\Delta_q\propto\exp(-\hbar\omega_p/E_{J1})$.  Thus, it would be unrealistic to expect a large scale processor involving a multitude of such devices to yield from even the best superconducting fabrication facility.  Moreover, implementation of AQO requires the ability to actively tune $\Delta_q$ from being the dominant energy scale in the qubit to being essentially negligible during the course of a computation.  Thus the single junction rf-SQUID flux qubit is of limited practical utility and has passed out of favor as a prototype qubit.

The next step in the evolution of the single junction flux qubit and related variants was the compound Josephson junction (CJJ) rf-SQUID, as depicted in Fig.~\ref{fig:rfssummary}b.  This device was first reported upon by Han, Lapointe and Lukens \cite{CJJ} and was the first type of flux qubit to display signatures of quantum superposition of macroscopic states \cite{LukensSuperposition}.  The CJJ rf-SQUID has been used by other research groups\cite{ItaliansCJJ,NiftyItalianPaper,MoreNiftyItalianPaper} and a related 4-Josephson junction device has been proposed \cite{3JJfluxqubits,MooijMore3JJFluxQubits}.  The CJJ rf-SQUID flux qubit and related variants have reappeared in a gradiometric configuration in more recent history \cite{HOMRT,IBM,Delft}.  Here, the single junction of Fig.~\ref{fig:rfssummary}a has been replaced by a flux biased dc-SQUID of inductance $L_{\text{cjj}}$ that allows one to tune the critical current of the rf-SQUID in-situ.  Let the applied flux threading this structure be denoted by $\Phicjjx$.  It is shown in Appendix A that the Hamiltonian for this system can be written as
\begin{subequations}
\begin{equation}
\label{eqn:2JHeff}
{\cal H}=\sum_n\left[\frac{Q_n^2}{2C_n}+U_n\frac{\left(\varphi_n-\varphi_n^x\right)^2}{2}\right]-U_q\beta_{\text{eff}}\cos\left(\varphi_q-\varphi_q^0\right) \; ,
\end{equation}

\noindent where the sum is over $n\in\left\{q,\text{cjj}\right\}$, $C_q\equiv C_1+C_2$, $1/C_{\text{cjj}}\equiv 1/C_1+1/C_2$ and $L_q\equiv L_{\text{body}}+L_{\text{cjj}}/4$.  The 2-dimensional potential energy in Hamiltonian (\ref{eqn:2JHeff}) is characterized by
\begin{equation}
\label{eqn:2JBeff}
\beta_{\text{eff}}=\beta_+\cos\left(\frac{\varphi_{\text{cjj}}}{2}\right)\sqrt{1+\left[\frac{\beta_-}{\beta_+}\tan(\varphi_{\text{cjj}}/2)\right]^2} \; ;
\end{equation}
\vspace{-12pt}
\begin{equation}
\label{eqn:2JOffset}
\varphi_q^0\equiv 2\pi\frac{\Phi_q^0}{\Phi_0} =-\arctan\left(\frac{\beta_-}{\beta_+}\tan\left(\varphi_{\text{cjj}}/2\right)\right) \; ;
\end{equation}
\vspace{-12pt}
\begin{equation}
\label{eqn:2Jbetapm}
\beta_{\pm}\equiv 2\pi L_q\left(I_{1}\pm I_{2}\right)/\Phi_0 \; .
\end{equation}
\end{subequations}

\noindent Note that if $\cos\left(\varphi_{\text{cjj}}/2\right)<0$, then $\beta_{\text{eff}}<0$ in Hamiltonian (\ref{eqn:2JHeff}).  This feature provides a natural means of shifting the qubit degeneracy point from $\varphi_q^x=\pi$, as in the single junction rf-SQUID case, to $\varphi_q^x\approx 0$.  It has been assumed in all that follows that this $\pi$-shifted mode of operation of the CCJ rf-SQUID has been invoked.

Hamiltonian (\ref{eqn:2JHeff}) is similar to that of a single junction rf-SQUID modulo the presence of a $\varphi_{\text{cjj}}$-dependent tunnel barrier through $\beta_{\text{eff}}$ and an effective critical current $I_q^c\equiv I_1+I_2$.
For $L_{\text{cjj}}/L_q\ll 1$ it is reasonable to assume that $\varphi_{\text{cjj}}\approx 2\pi\Phicjjx/\Phi_0$.  Consequently, the CJJ rf-SQUID facilitates in-situ tuning of the tunneling energy through $\Phicjjx$.  While this is clearly desirable, one does pay for the additional flexibility by adding more complexity to the rf-SQUID design and thus more potential room for fabrication variations.  The minimum achievable barrier height is ultimately limited by any so called {\it junction asymmetry} which leads to finite $\beta_{-}$.  In practice, for $\beta_-/\beta_+=(I_{1}-I_{2})/(I_{1}+I_{2})\lesssim0.05$, this effect is of little concern.  However, a more insidious effect of junction asymmetry can be seen via the change of variables $\varphi_q-\varphi_q^0\rightarrow\varphi_q$ in Eq.~(\ref{eqn:2JHeff}), namely an apparent $\Phicjjx$-dependent flux offset: $\Phiqx\rightarrow\Phiqx-\Phi_q^0(\Phicjjx)$.  If the purpose of the CJJ is to simply allow the experimentalist to target a particular $\Delta_q$, then the presence of $\Phi_q^0(\Phicjjx)$ can be readily compensated via the application of a static flux offset.  On the other hand, any mode of operation that explicitly requires altering $\Delta_q$ during the course of a quantum computation \cite{IBM,Oliver,Kaminsky,Aharonov,NiftyItalianPaper,MoreNiftyItalianPaper} would also require active compensation for what amounts to a nonlinear crosstalk from $\Phicjjx$ to $\Phiqx$.  While it may be possible to approximate this effect as a linear crosstalk over a small range of $\Phicjjx$ if the junction asymmetry is small, one would nonetheless need to implement precise {\it time-dependent} flux bias compensation to utilize the CJJ rf-SQUID as a flux qubit in any quantum computation scheme.  While this may be feasible in laboratory scale systems, it is by no means desirable nor practical on a large scale quantum information processor.

A second problem with the CJJ rf-SQUID flux qubit is that one cannot homogenize the qubit parameters $\iqp$ and $\Delta_q$ between a multitude of such devices that possess different $\beta_{\pm}$ over a broad range of $\Phicjjx$.  While one can accomplish this task to a limited degree in a perturbative manner about carefully chosen CJJ biases for each qubit \cite{synchronization}, the equivalence of $\iqp$ and $\Delta_q$ between those qubits will be approximate at best.  Therefore, the CJJ rf-SQUID does not provide a convenient means of accommodating fabrication variations between multiple flux qubits in a large scale processor.

Given that the CJJ rf-SQUID provides additional flexibility at a cost, it is by no means obvious that one can design a better rf-SQUID flux qubit by adding even more junctions.  Specifically, it is desirable to have a device whose imperfections can be mitigated purely by the application of {\it time-independent} compensation signals.  The novel rf-SQUID topology shown in Fig.~\ref{fig:rfssummary}c, hereafter referred to as the compound-compound Josephson junction (CCJJ) rf-SQUID, satisfies this latter constraint.  Here, each junction of the CJJ in Fig.~\ref{fig:rfssummary}b has been replaced by a dc-SQUID, which will be referred to as left ($L$) and right ($R$) minor loops, and will be subjected to external flux biases $\Phi_L^x$ and $\Phi_R^x$, respectively.  The role of the CJJ loop in  Fig.~\ref{fig:rfssummary}b is now played by the CCJJ loop of inductance $L_{\text{ccjj}}$ which will be subjected to an external flux bias $\Phiccjjx$.  It is shown in Appendix B that if one chooses {\it static} values of $\Phi_L^x$ and $\Phi_R^x$ such that the net critical currents of the minor loops are equal, then it can be described by an effective two-dimensional Hamiltonian of the form
\begin{subequations}
\begin{equation}
\label{eqn:4JHeff}
{\cal H}=\sum_n\left[\frac{Q_n^2}{2C_n}+U_n\frac{\left(\varphi_n-\varphi_n^x\right)^2}{2}\right]-U_q\beta_{\text{eff}}\cos\left(\varphi_q-\varphi_q^0\right) \; ,
\end{equation}

\noindent where the sum is over $n\in\left\{q,\text{ccjj}\right\}$, $C_q\equiv C_1+C_2+C_3+C_4$, $1/C_{\text{ccjj}}\equiv 1/(C_1+C_2)+1/(C_3+C_4)$ and $L_q\equiv L_{\text{body}}+L_{\text{ccjj}}/4$.  The effective 2-dimensional potential energy in Hamiltonian (\ref{eqn:4JHeff}) is characterized by
\begin{equation}
\label{eqn:4JBeffbalanced}
\beta_{\text{eff}}=\beta_+(\Philx,\Phirx)\cos\left(\frac{\varphi_{\text{ccjj}}-\varphi^0_{\text{ccjj}}}{2}\right) \;\; ,
\end{equation}

\noindent where $\beta_+(\Philx,\Phirx)=2\pi L_q I_q^c(\Philx,\Phirx)/\Phi_0$ with 
\begin{displaymath}
I_q^c(\Philx,\Phirx)\equiv (I_1+I_2)\cos\left(\frac{\pi\Philx}{\Phi_0}\right)+(I_3+I_4)\cos\left(\frac{\pi\Phirx}{\Phi_0}\right) \; .  
\end{displaymath}

\noindent Given an appropriate choice of $\Philx$ and $\Phirx$, the $q$ and ccjj loops will possess apparent flux offsets of the form
\begin{equation}
\label{eqn:qoffset}
\Phi_q^0=\frac{\Phi_0\varphi_q^0}{2\pi}=\frac{\Phi_{L}^0+\Phi_{R}^0}{2}\; ;
\end{equation}
\vspace{-12pt}
\begin{equation}
\label{eqn:ccjjoffset}
\Phi^0_{\text{ccjj}}=\frac{\Phi_0\varphi^0_{\text{ccjj}}}{2\pi}=\Phi_{L}^0-\Phi_{R}^0\; ,
\end{equation}
\end{subequations}

\noindent where $\Phi_{L(R)}^0$  is given by Eq.~(\ref{eqn:4JMinorOffset}), which is purely a function of $\Phi^x_{L(R)}$ and junction critical currents.  As such, the apparent flux offsets are {\it independent} of $\Phiccjjx$.  Under such conditions, we deem the CCJJ to be {\it balanced}.  Given that the intended mode of operation is to hold $\Phi_L^x$ and $\Phi_R^x$ constant, then the offset phases  $\varphi_L^0$ and $\varphi_R^0$ will also be constant.  The result is that Hamiltonian (\ref{eqn:4JHeff}) for the CCJJ rf-SQUID becomes homologous to that of an ideal CJJ rf-SQUID [$\beta_-=0$ in Eqs.~(\ref{eqn:2JBeff}) and (\ref{eqn:2JOffset})] with apparent {\it static} flux offsets.  Such static offsets can readily be calibrated and compensated for in-situ using either analog control lines or on-chip programmable flux sources \cite{PCC}.  For typical device parameters and junction variability on the order of a few percent, these offsets will be $\sim 1\rightarrow 10\,$m$\Phi_0$.  Equations \ref{eqn:4JHeff}-\ref{eqn:ccjjoffset} with $\Phi_q^0=\Phi^0_{\text{ccjj}}=0$ will be referred to hereafter as the ideal CCJJ rf-SQUID model.

The second advantage of the CCJJ rf-SQUID is that one can readily accommodate for variations in critical current between multiple flux qubits.  Note that in Eq.~(\ref{eqn:4JBeffbalanced}) that the maximum height of the tunnel barrier is governed by $\beta_+(\Philx,\Phirx)\equiv\beta_L(\Philx)+\beta_R(\Phirx)$, where $\beta_{L(R)}$ is given by Eq.~(\ref{eqn:4JMinorOffset}).  One is free to choose any pair of $(\Philx,\Phirx)$ such that $\beta_L(\Philx)=\beta_R(\Phirx)$, as dictated by Eq.~(\ref{eqn:balancedapprox}).  Consequently, $\beta_+=2\beta_R(\Phirx)$ in Eq.~(\ref{eqn:4JBeffbalanced}).  One can then choose $\Phirx$, which then dictates $\Philx$, so as to homogenize $\beta_+$ between multiple flux qubits.  The results is a set of nominally uniform flux qubits where the particular choice of $(\Philx,\Phirx)$ for each qubit merely results in unique static flux offsets $\Phi^0_q$ and $\Phi^0_{\text{ccjj}}$ for each device.

To summarize up to this point, the CCJJ rf-SQUID is robust against Josephson junction fabrication variations both within an individual rf-SQUID and between a plurality of such devices.  The variations can be effectively tuned out purely through the application of {\it static} flux biases, which is of considerable advantage when envisioning the implementation of large scale quantum information processors that use flux qubits.

\subsection{$L$-tuner}

The purpose of the CCJJ structure was to provide a means of coming to terms with fabrication variations in Josephson junctions both within individual flux qubits and between sets of such devices.  However, junctions are not the only key parameter that may vary between devices, nor are fabrication variations responsible for all of the potential variation.  In particular, it has been experimentally demonstrated that the inductance of a qubit $L_q$ that is connected to other qubits via tunable mutual inductances is a function of the coupler configuration \cite{cjjcoupler}.  Let the bare inductance of the qubit in the presence of no couplers be represented by $L_q^0$ and the mutual inductance between the qubit and coupler $i$ be represented by $M_{\text{co},i}$.  If the coupler possesses a first order susceptibility $\chi_i$, as defined in Ref.~\onlinecite{cjjcoupler}, then the net inductance of the qubit can be expressed as 
\begin{equation}
\label{eqn:LqNoTuner}
L_q=L_q^0-\sum_{i}M^2_{\text{co},i}\chi_i\; .        
\end{equation}

\noindent Given that qubit properties such as $\Delta_q$ can be exponentially sensitive to variations in $L_q$, then it is undesirable to have variations in $L_q$ between multiple flux qubits or to have $L_q$ change during operation.  This could have a deleterious impact upon AQO in which it is typically assumed that all qubits are identical and they are intended to be annealed in unison \cite{AQC}.  From the perspective of GMQC, one could very well attempt to compensate for such effects in a CJJ or CCJJ rf-SQUID flux qubit by adjusting the tunnel barrier height to hold $\Delta_q$ constant, but doing so alters $\iqp$, which then alters the coupling of the qubit to radiative sources, thus demanding further compensation.  Consequently, it also makes sense from the perspective of GMQC that one find a means of rendering $L_q$ uniform between multiple qubits and insensitive to the settings of inductive coupling elements.

\begin{figure}
\includegraphics[width=2.5in]{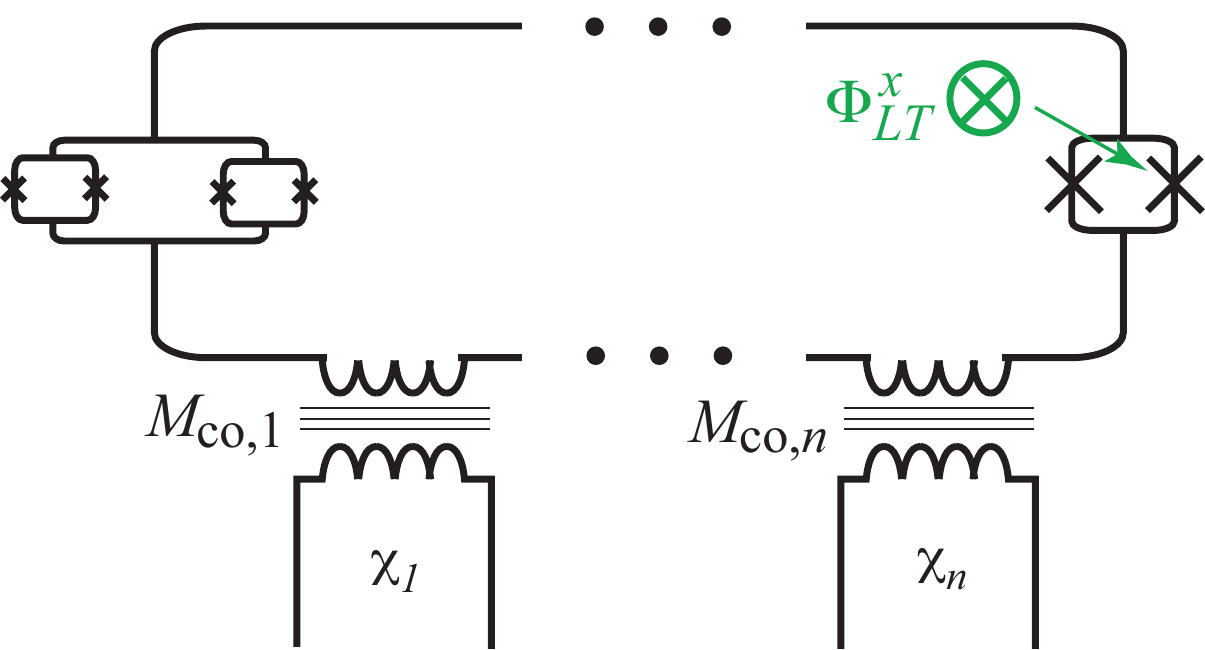}
\caption{\label{fig:LTuner}  (color online) A CCJJ rf-SQUID with $L$-tuner connected to multiple tunable inductive couplers via transformers with mutual inductances $M_{\text{co},i}$ and possessing susceptibilities $\chi_i$.  The $L$-tuner is controlled via the external flux bias $\Phi^x_{LT}$}
\end{figure}

In order to compensate for variations in $L_q$, we have inserted a tunable Josephson inductance \cite{vanDuzer} into the CCJJ rf-SQUID body, as depicted in Fig.~\ref{fig:LTuner}.   We refer to this element as an inductance ($L$)-tuner.  This relatively simple element comprises a dc-SQUID whose critical current vastly exceeds that of the CCJJ structure, thus ensuring negligible phase drop across the $L$-tuner.  Assuming that the inductance of the $L$-tuner wiring is negligible, the $L$-tuner modifies Eq.~(\ref{eqn:LqNoTuner}) in the following manner:
\begin{equation}
\label{eqn:LqWithTuner}
L_q=L_q^0-\sum_{i}M^2_{\text{co},i}\chi_i + \frac{L_{J0}}{\cos(\pi\Phi^x_{LT}/\Phi_0)}\; ,
\end{equation}

\noindent where $L_{J0}\equiv\Phi_0/2\pi I^c_{LT}$, $I^c_{LT}$ is the net critical current of the two junctions in the $L$-tuner and $\Phi^x_{LT}$ is an externally applied flux bias threading the $L$-tuner loop.  For modest flux biases such that $I^c_{LT}\cos(\pi\Phi^x_{LT}/\Phi_0)\gg I_q^c$, Eq.~(\ref{eqn:LqWithTuner}) is a reliable model of the physics of the $L$-tuner.

Given that the $L$-tuner is only capable of augmenting $L_q$, one can only choose to target $L_q>L_q^0-\sum_iM_{\text{co},i}^2\chi^{\text{AFM}}_i+L_{J0}$, where $\chi^{\text{AFM}}_i$ is the maximum antiferromagnetic (AFM) susceptibility of inter-qubit coupler $i$.  In practice, we choose to restrict operation of the couplers to the range $-\chi_i^{\text{AFM}}<\chi_i<\chi_i^{\text{AFM}}$ such that the maximum qubit inductance that will be encountered is $L_q>L_q^0+\sum_iM_{\text{co},i}^2\chi^{\text{AFM}}_i+L_{J0}$.  We then choose to prebias $\Phi^x_{\text{LT}}$ for each qubit to match the maximum realized $L_q\equiv L^{\text{max}}_q$ amongst a set of flux qubits.  Thereafter, one can hold $L_q=L^{\text{max}}_q$ as couplers are adjusted by inverting Eq.~(\ref{eqn:LqWithTuner}) to solve for an appropriate value of $\Phi^x_{LT}$.  Thus, the $L$-tuner provides a ready means of compensating for small variations in $L_q$ between flux qubits and to hold $L_q$ constant as inductive inter-qubit coupling elements are adjusted.

\section{Device Architecture, Fabrication and Readout Operation}

\begin{figure}
\includegraphics[width=3.25in]{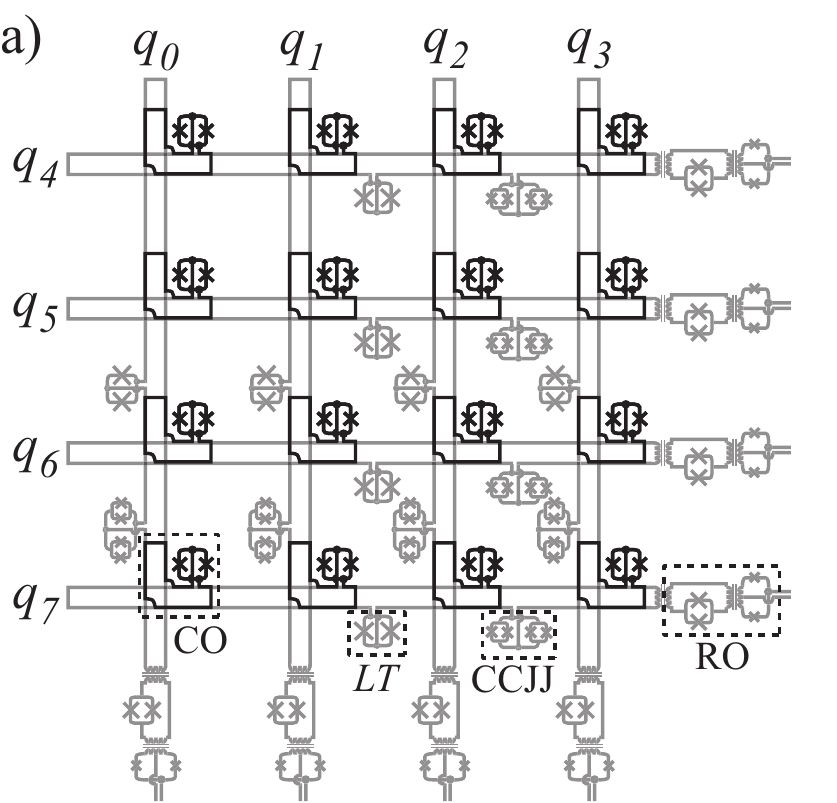} \\
\includegraphics[width=3.25in]{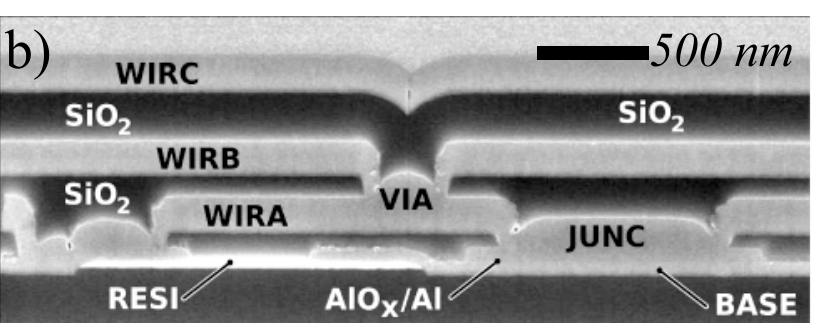} \\
\includegraphics[width=3.25in]{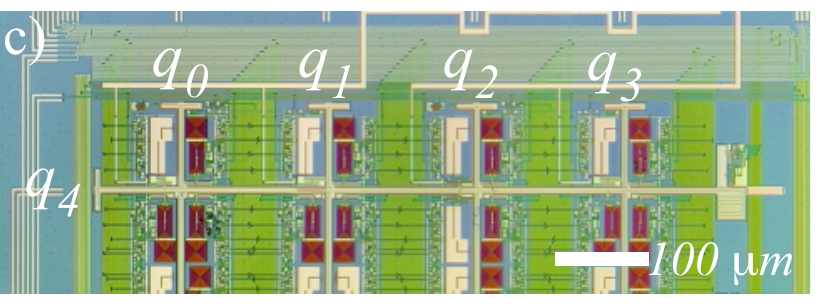}
\caption{\label{fig:architecture}  (color online) a)  High level schematic of the analog devices on the device reported upon herein.  Qubits are represented as light grey elongated objects and denoted as $q_0\ldots q_7$.  One representative readout (RO), CCJJ and $L$-tuner ($LT$) each have been indicated in dashed boxes.   Couplers (CO) are represented as dark objects located at the intersections of the qubit bodies.  b)  SEM of a cross-section of the fabrication profile.  Metal layers denoted as BASE, WIRA, WIRB and WIRC.  Insulating layers labeled as SiO$_2$.   Topmost insulator has not been planarized in this test structure, but is planarized in the full circuit process.  An example via (VIA), Josephson junction (JUNC, AlO$_x$/Al) and resistor (RESI) are also indicated.  c)  Optical image of a portion of a device completed up to WIRB.  Portions of qubits $q_0\ldots q_3$ and the entirety of $q_4$ are visible.}
\end{figure}

To test the CCJJ rf-SQUID flux qubit, we fabricated a circuit containing 8 such devices with pairwise interactions mediated by a network of 16 in-situ tunable CJJ rf-SQUID inter-qubit couplers \cite{cjjcoupler}.  Each qubit was also coupled to its own dedicated quantum flux parametron (QFP)-enabled readout \cite{QFP}.  A high level schematic of the device architecture is shown in Fig.~\ref{fig:architecture}a.  External flux biases were provided to target devices using a sparse combination of analog current bias lines to facilitate device calibration and an array of single flux quantum (SFQ) based on-chip programmable control circuitry (PCC) \cite{PCC}.  

The device was fabricated from an oxidized Si wafer with Nb/Al/Al$_2$O$_3$/Nb trilayer junctions and four Nb wiring layers separated by planarized plasma enhanced chemical vapor deposited SiO$_{2}$.  A scanning electron micrograph of the process cross-section is shown in Fig.~\ref{fig:architecture}b.  The Nb metal layers have been labeled as BASE, WIRA, WIRB and WIRC.  The flux qubit wiring was primarily located in WIRB and consisted of $2\,\mu$m wide leads arranged as an approximately $900\,\mu$m long differential microstrip located $200\,$nm above a groundplane in WIRA.  CJJ rf-SQUID coupler wiring was primarily located in WIRC, stacked on top of the qubit wiring to provide inductive coupling.  PCC flux storage loops were implemented as stacked spirals of 13-20 turns of $0.25\,\mu$m wide wiring with $0.25\,\mu$m separation in BASE and WIRA (WIRB).  Stored flux was picked up by one-turn washers in WIRB (WIRA) and fed
into transformers for flux-biasing devices.  External control lines were mostly located in
BASE and WIRA.  All of these control elements resided below a groundplane in WIRC.  The groundplane under the qubits and over the PCC/external control lines were electrically connected using extended vias in WIRB so as to form a  nearly continuous superconducting shield between the analog devices on top and the bias circuitry below.  To provide biases to target devices with minimal parasitic crosstalk, transformers for biasing qubits, couplers, QFPs and dc-SQUIDs using bias lines and/or PCC elements were enclosed in superconducting boxes with BASE and WIRC forming the top and bottom, respectively, and vertical walls formed by extended vias in WIRA and WIRB.  Minimal sized openings were placed in the vertical walls through which the bias and target device wiring passed at opposing ends of each box.  

An optical image of a portion of a device completed up to WIRB is shown in Fig.~\ref{fig:architecture}c.  Qubits are visible as elongated objects,  WIRB PCC spirals are visible as dark rectangles and WIRB washers are visible as light rectangles with slits.  Note that the extrema of the CCJJ rf-SQUID qubits are terminated in unused transformers.  These latter elements allow this 8-qubit unit cell to be tiled in a larger device with additional inter-qubit CJJ couplers providing the connections between unit cells.

\begin{figure}
\includegraphics[width=3.25in]{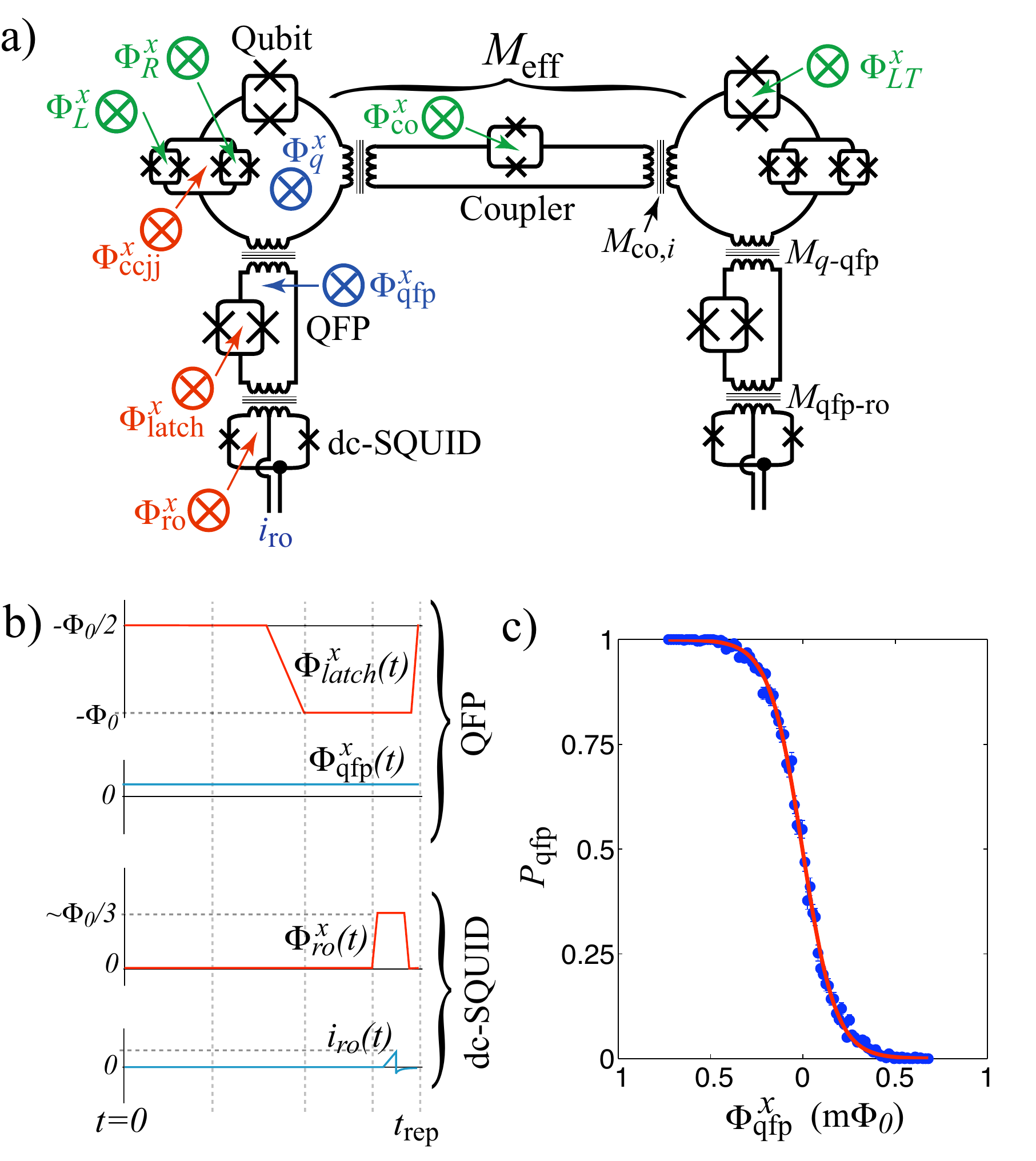}
\caption{\label{fig:unipolarannealing}  (color online) a)  Schematic representation of a portion of the circuit reported upon herein.  Canonical representations of all externally controlled flux biases $\Phi_{\alpha}^x$, readout current bias $i_{ro}$ and key mutual inductances $M_{\alpha}$ are indicated.  b)  Depiction of latching readout waveform sequence.  c)  Example QFP state population measurement as a function of the dc level $\Phi^x_{\text{qfp}}$ with no qubit signal.  Data have been fit to Eq.~(\ref{eqn:transition}).}
\end{figure}

We have studied the properties of all 8 CCJJ rf-SQUID flux qubits on this chip in detail and report upon one such device herein.  To clearly establish the lingua franca of our work, we have depicted a portion of the multi-qubit circuit in Fig.~\ref{fig:unipolarannealing}a.  Canonical representations of the external flux biases needed to operate a qubit, a coupler and a QFP-enabled readout are labeled on the diagram.  The fluxes $\Phi_L^x$, $\Phi_R^x$, $\Phi^x_{LT}$ and $\Phi_{\text{co}}^x$ were only ever subjected to dc levels in our experiments that were controlled by the PCC.  The remaining fluxes and readout current biases were driven by a custom-built 128 channel room temperature current source.  The mutual inductance between qubit and QFP ($M_{q-\text{qfp}}$), between QFP and dc-SQUID ($M_{\text{qfp-ro}}$), qubit and coupler ($M_{\text{co},i}$) and $\Phicox$-dependent inter-qubit mutual inductance ($M_{\text{eff}}$) have also been indicated.  Further details concerning cryogenics, magnetic shielding and signal filtering have been discussed in previous publications \cite{LOMRT,PCC,QFP,cjjcoupler}.

Since much of what follows depends upon a clear understanding of our novel QFP-enabled readout mechanism, we present a brief review of its operation herein.  The flux and readout current waveform sequence involved in a single-shot readout is depicted in Fig.~\ref{fig:unipolarannealing}b.  Much like the CJJ qubit \cite{LOMRT}, the QFP can be adiabatically {\it annealed} from a state with a monostable potential ($\Philatchx=-\Phi_0/2$) to a state with a bistable potential ($\Philatchx=-\Phi_0$) that supports two counter-circulating persistent current states.  The matter of which persistent current state prevails at the end of an annealing cycle depends upon the sum of $\Phiqfpx$ and any signal from the qubit mediated via $M_{q-\text{qfp}}$.  The state of the QFP is then determined with high fidelity using a synchronized flux pulse and current bias ramp applied to the dc-SQUID.  The readout process was typically completed within a repetition time $t_{\text{rep}}<50\,\mu$s.

An example trace of the population of one of the QFP persistent current states $P_{\text{qfp}}$ versus $\Phiqfpx$, obtained using the latching sequence depicted in Fig.~\ref{fig:unipolarannealing}b, is shown in Fig.~\ref{fig:unipolarannealing}c.  This trace was obtained with the qubit potential held monostable ($\Phiccjjx=-\Phi_0/2$) such that it presented minimal flux to the QFP and would therefore not influence $P_{\text{qfp}}$.  The data have been fit to the phenomenological form
\begin{equation}
\label{eqn:transition}
P_{\text{qfp}}=\frac{1}{2}\left[1-\tanh\left(\frac{\Phiqfpx-\Phi^0_{\text{qfp}}}{w}\right)\right]
\end{equation}

\noindent  with width $w\sim 0.18\,$m$\Phi_0$ for the trace shown therein.  When biased with constant $\Phiqfpx=\Phi^0_{\text{qfp}}$, which we refer to as the QFP degeneracy point, this transition in the population statistics can be used as a highly nonlinear flux amplifier for sensing the state of the qubit.  Given that $M_{q-\text{qfp}}=6.28\pm0.01\,$pH for the devices reported upon herein and that typical qubit persistent currents in the presence of negligible tunneling $\iqp\gtrsim 1\,\mu$A, then the net flux presented by a qubit was $2M_{q-\text{qfp}}\iqp\gtrsim 6\,$m$\Phi_0$, which far exceeded $w$.  By this means one can achieve the very high qubit state readout fidelity reported in Ref.~\onlinecite{QFP}.  On the other hand, the QFP can be used as a linearized flux sensor by engaging $\Phiqfpx$ in a feedback loop and actively tracking $\Phi^0_{\text{qfp}}$.  This latter mode of operation has been used extensively in obtaining many of the results presented herein.

\section{CCJJ rf-SQUID Characterization}

The purpose of this section is to present measurements that characterize the CCJJ, $L$-tuner and capacitance of a CCJJ rf-SQUID.  All measurements shown herein have been made with a set of standard bias conditions given by $\Philx=98.4\,$m$\Phi_0$, $\Phirx=-89.3\,$m$\Phi_0$, $\Phi^x_{\text{LT}}=0.344\,\Phi_0$ and all inter-qubit couplers tuned to provide $M_{\text{eff}}=0$, unless indicated otherwise.  The logic behind this particular choice of bias conditions will be explained in what follows.
This section will begin with a description of the experimental methods for extracting $L_q$ and $I_q^c$ from persistent current measurements.  Thereafter, data that demonstrate the performance of the CCJJ and $L$-tuner will be presented.  Finally, this section will conclude with the determination of $C_q$ from macroscopic resonant tunneling data.

\subsection{High Precision Persistent Current Measurements}  

The most direct means of obtaining information regarding a CCJJ rf-SQUID is to measure the persistent current $\iqp$ as a function of $\Phiccjjx$.  A reasonable first approach to measuring this quantity would be to sequentially prepare the qubit in one of its persistent current states and then the other, and use the QFP in feedback mode to measure the difference in flux sensed by the QFP, which equals $2M_{q-\text{qfp}}\iqp$.  A fundamental problem with this approach is that it is sensitive to low frequency (LF) flux noise \cite{1OverF}, which can alter the background flux experienced by the QFP between the sequential measurements.  For a typical measurement with our apparatus, the act of locating a single QFP degeneracy point to within $20\,\mu\Phi_0$ takes on the order of $1\,$s, which means that two sequential measurements would only be immune to flux noise below $0.5\,$Hz.  We have devised a LF flux noise rejection scheme that takes advantage of the fact that such noise will generate a correlated shift in the apparent degeneracy points if the sequential preparations of the qubit can be interleaved with single-shot measurements that are performed in rapid succession.  If these measurements  are performed with repetition time $t_{\text{rep}}\sim 1\,$ms, then the measurements will be immune to flux noise below $\sim 1\,$kHz.

\begin{figure}
\includegraphics[width=3.25in]{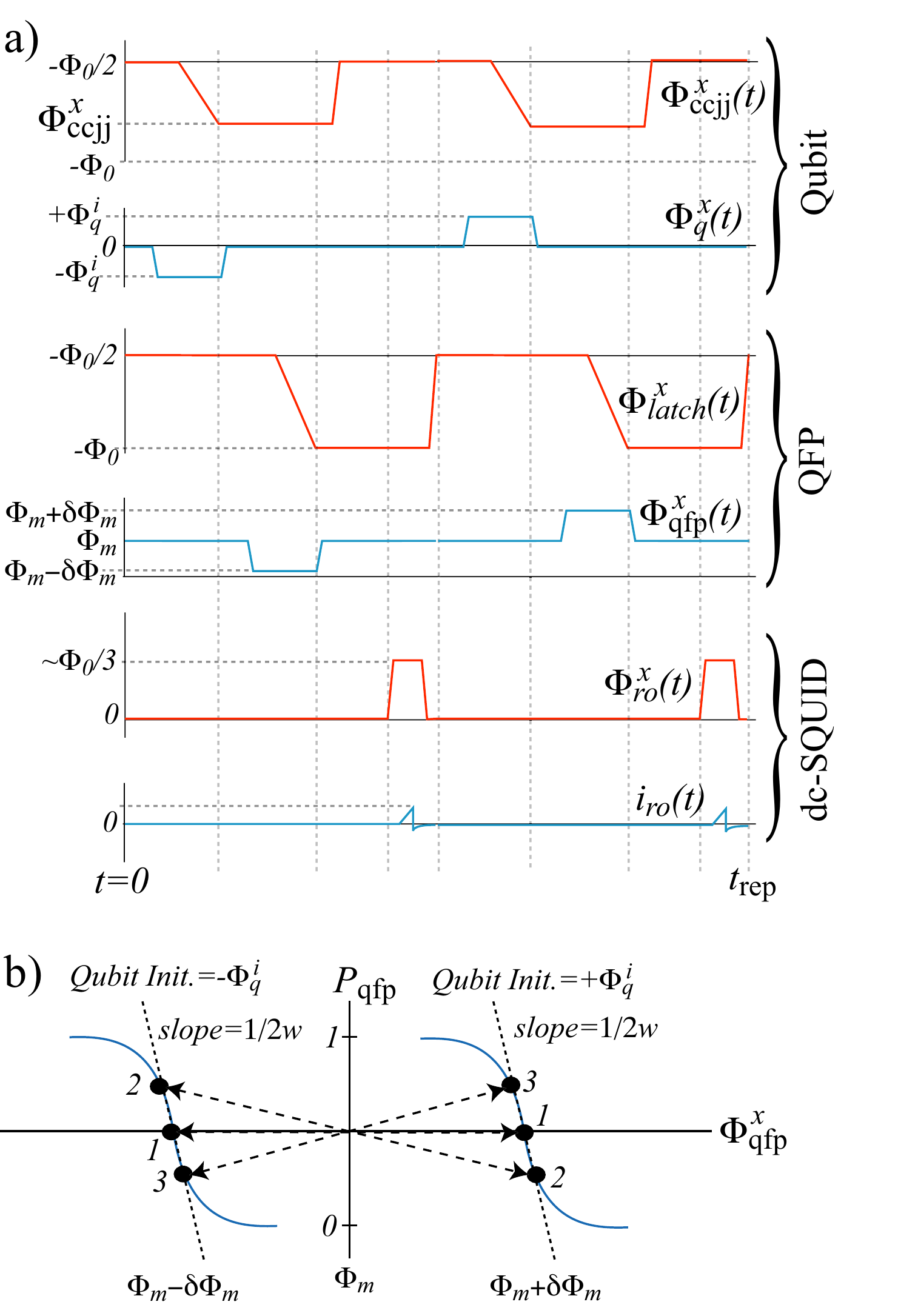}
\caption{\label{fig:iqplockin} (color online)  a)  Low frequency flux noise rejecting qubit persistent current measurement sequence.  Waveforms shown are appropriate for measuring $\left|I_q^p\left(\Phiccjjx\right)\right|$ for $-\Phi_0\leq\Phiccjjx\leq 0$.  The $\Phiccjjx$ waveform can be offset by integer $\Phi_0$ to measure the periodic behavior of this quantity.  Typical repetition time is $t_{\text{rep}}\sim 1\,$ms. b)  Depiction of QFP transition and correlated changes in QFP population statistics for the two different qubit initializations.}
\end{figure}

A depiction of the LF flux noise rejecting persistent current measurement sequence is shown in Fig.~\ref{fig:iqplockin}a.  The waveforms comprise two concatenated blocks of sequential annealing of the qubit to a target $\Phiccjjx$ in the presence of an alternating polarizing flux bias $\pm\Phi_q^i$ followed by latching and single-shot readout of the QFP.  The QFP flux bias is engaged in a differential feedback mode in which it is pulsed in alternating directions by an amount $\delta\Phi_m$ about a mean level $\Phi_m$.  The two single-shot measurements yield binary results for the QFP state and the {\it difference} between the two binary results is recorded.  Gathering a statistically large number of such differential measurements then yields a differential population measurement $\delta P_{\text{qfp}}$.  Conceptually, the measurement works in the manner depicted in Fig.~\ref{fig:iqplockin}b: the two different initializations of the qubit move the QFP degeneracy point to some unknown levels $\Phi_m^0\pm\delta\Phi_m^0$, where $\Phi_m^0$ represents the true mean of the degeneracy points at any given instant in time and $2\delta\Phi_m^0$ is the true difference in degeneracy points that is independent of time.  Focusing on flux biases that are close to the degeneracy point, one can linearize Eq.~(\ref{eqn:transition}):
\begin{equation}
\label{eqn:transitionlinear}
P_{\text{qfp},\pm}\approx\frac{1}{2}+\frac{1}{2w}\left[\Phiqfpx-\left(\Phi_m^0\pm\delta\Phi_m^0\right)\right] \; .
\end{equation}

\noindent Assuming that the rms LF flux noise $\Phi_n\ll w$ and that one has reasonable initial guesses for $\Phi_m^0\pm\delta\Phi_m^0$, then the use of the linear approximation should be justified.  Applying $\Phiqfpx=\Phi_m\pm\delta\Phi_m$ and sufficient repetitions of the waveform pattern shown in Fig.~\ref{fig:iqplockin}a, the differential population will then be of the form
\begin{equation}
\label{eqn:diffpop}
\delta P_{\text{qfp}}=P_{\text{qfp},+}-P_{\text{qfp},-}=\frac{1}{w}\left[\delta\Phi_m+\delta\Phi_m^0\right]\; ,
\end{equation}

\noindent which is {\it independent} of $\Phi_m$ and $\Phi_m^0$.  Note that the above expression contains only two independent variables, $w$ and $\delta\Phi_m^0$, and that $\delta P_{\text{qfp}}$ is purely a linear function of $\delta\Phi_m$.  By sampling at three values of $\delta\Phi_m$, as depicted by the pairs of numbered points in Fig.~\ref{fig:iqplockin}b, the independent variables in Eq.~(\ref{eqn:diffpop}) will be overconstrained, thus readily yielding $\delta\Phi_m^0$.  One can then infer the qubit persistent current as follows:
\begin{equation}
\label{eqn:iqplockin}
\iqp=\frac{2\delta\Phi_m^0}{2M_{q-\text{qfp}}}=\frac{\delta\Phi_m^0}{M_{q-\text{qfp}}} \; .
\end{equation}

\begin{figure}[ht]
\includegraphics[width=3.25in]{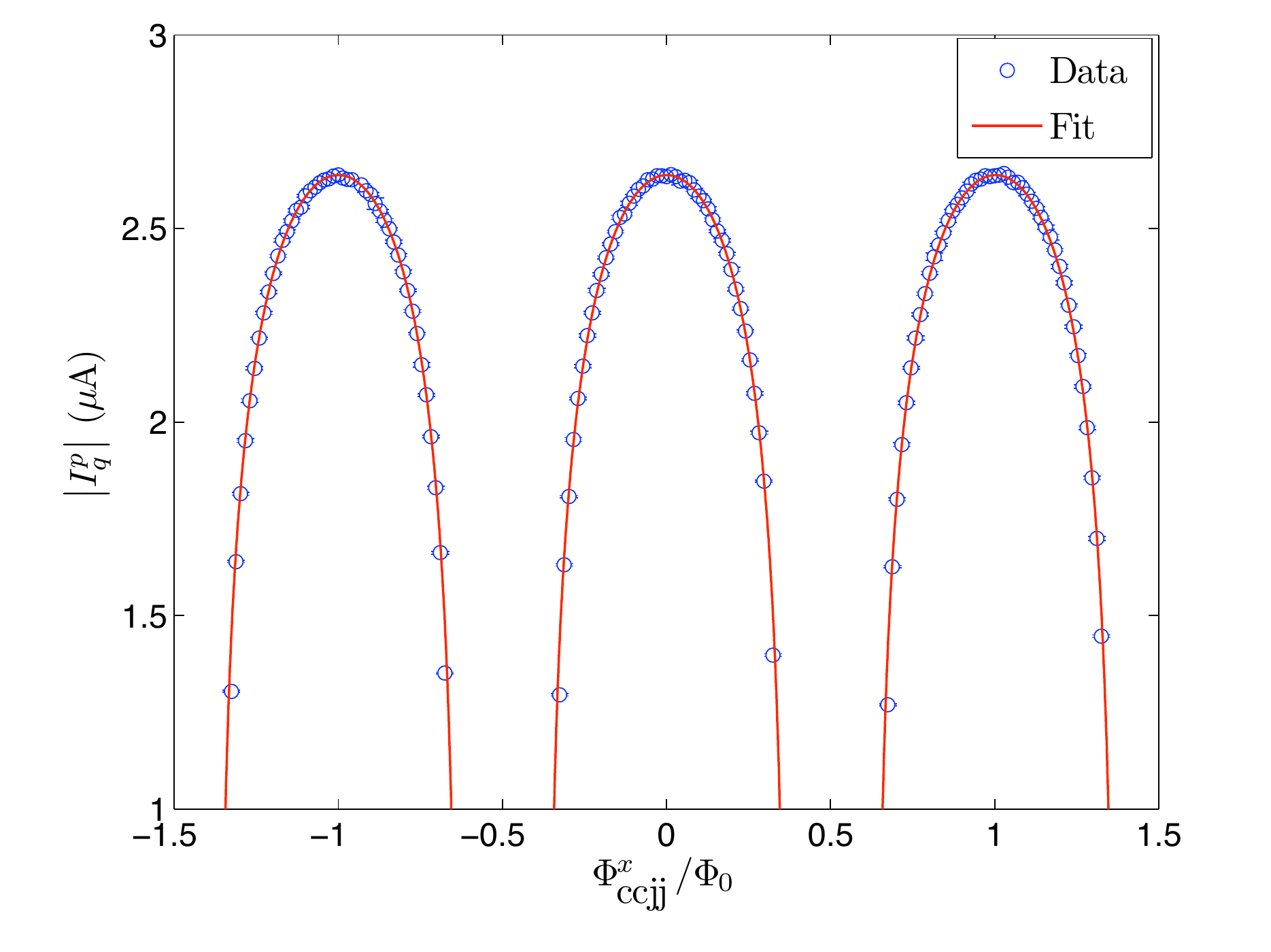}
\caption{\label{fig:Lq0extraction} (color online) Example measurements of $\left|I_q^p\left(\Phiccjjx\right)\right|$.}
\end{figure}

\noindent  Example measurements of $\iqp\left(\Phiccjjx\right)$ are shown in Fig.~\ref{fig:Lq0extraction}.  These data, for which $1.5\lesssim\left|\beta_{\text{eff}}\right|\lesssim 2.5$, have been fit to the ideal CCJJ rf-SQUID model by finding the value of $\varphi_q\equiv\varphi^{\text{min}}_q$ for which the potential in Eq.~(\ref{eqn:4JHeff}) is minimized:
\begin{equation}
\label{eqn:iqp2d}
\iqp=\frac{\Phi_0}{2\pi}\frac{\left|\varphi^{\text{min}}_q-\varphi_q^x\right|}{L_q} \;\; .
\end{equation}

\noindent The best fit shown in Fig.~\ref{fig:Lq0extraction} was obtained with $L_q=265.4 \pm 1.0\,$pH, $L_{\text{ccjj}}=26\pm 1\,$pH and $I_q^c=3.103\pm 0.003\,\mu$A.  For comparison, we had estimated $L_q=273\,$pH at the standard bias condition for $\Phi^x_{LT}$ and $L_{\text{ccjj}}=20\,$pH from design.  

In practice, we have found that the LF flux noise rejecting method of measuring $\iqp$ effectively eliminates any observable $1/f$ component in that measurement's noise power spectral density, to within statistical error.  Finally, it should be noted that the LF flux noise rejecting method is applicable to any measurement of a difference in flux sensed by a linearized detector.  In what follows herein, we have made liberal use of this technique to calibrate a variety of quantities in-situ using both QFPs and other qubits as flux detectors.

\subsection{CCJJ}

\begin{figure}[ht]
\includegraphics[width=3.25in]{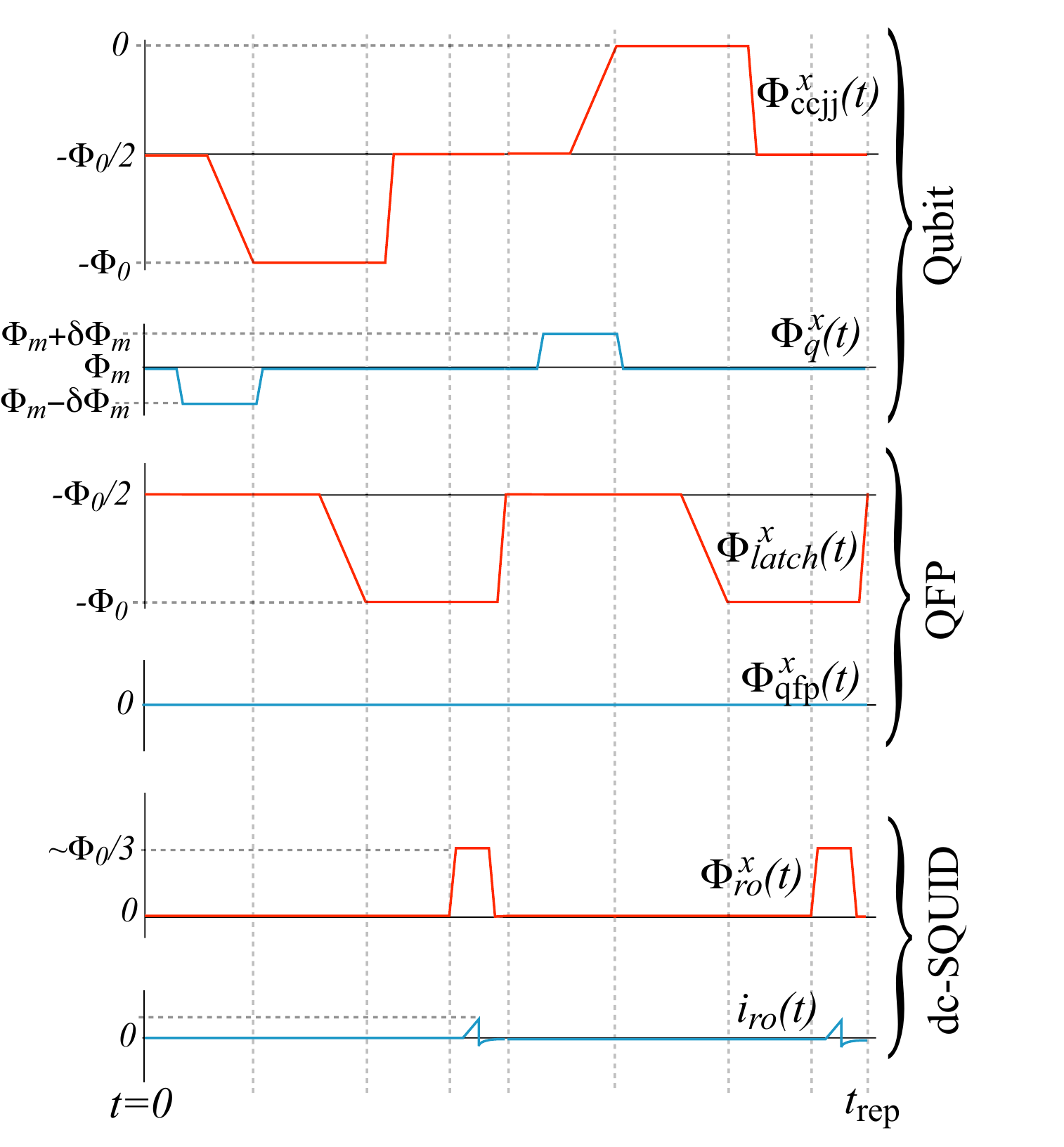}
\caption{\label{fig:bipolarlockin}  (color online) Schematic of low frequency noise rejecting differential qubit degeneracy point measurement sequence.  The qubit is annealed with a $\Phiccjjx$ signal of opposing polarity in the two frames and the qubit flux bias is controlled via feedback.}
\end{figure}

In this subsection, the CCJJ has been characterized as a function of $\Philx$ and $\Phirx$ with all other static flux biases set to the standard bias condition cited above.  Referring to Eq.~(\ref{eqn:4JQOffset}), it can be seen that the qubit degeneracy point $\Phi_q^0$ is a function of $\Phiccjjx$ through $\gamma_0$ if the CCJJ has not been balanced. To accentuate this functional dependence, one can anneal the CCJJ rf-SQUID with $\Phiccjjx$ waveforms of opposing polarity about a minimum in $\left|\beta_{\text{eff}}\right|$, as found at $\Phiccjjx=-\Phi_0/2$.  The expectation is that the {\it apparent} qubit degeneracy points will be antisymmetric about the mean given by setting $\gamma_0=0$ in Eq.~(\ref{eqn:4JQOffset}).  The waveform sequence for performing a differential qubit degeneracy point measurement is depicted in Fig.~\ref{fig:bipolarlockin}.  In this case, the QFP is used as a latching readout and the qubit acts as the linearized detector of its own apparent annealing polarization-dependent flux offset.  As with the $\iqp$ measurement described above, this LF flux noise rejecting procedure returns a {\it difference} in apparent flux sensed by the qubit and not the absolute flux offsets.  

\begin{figure}[ht]
\includegraphics[width=3.25in]{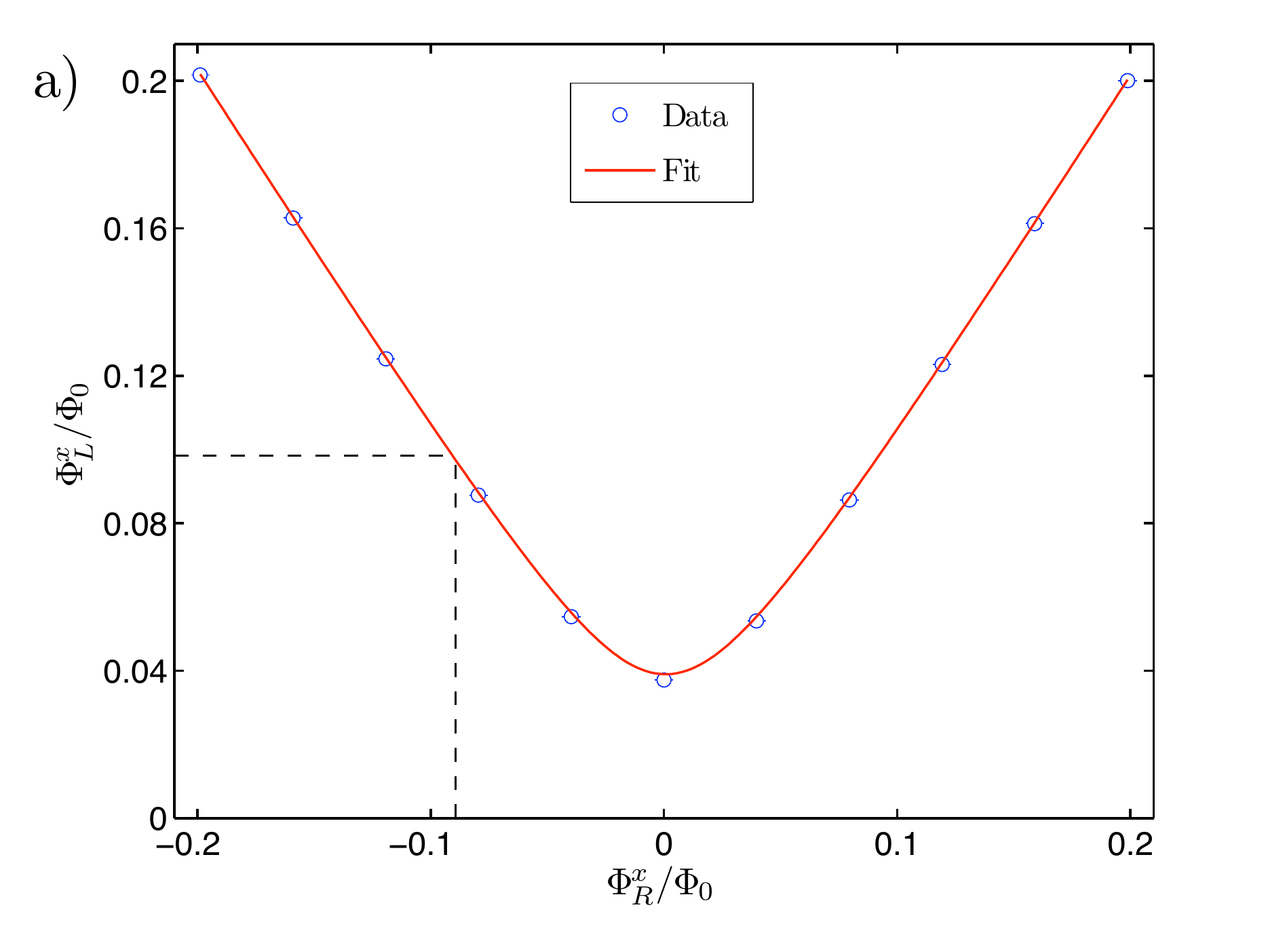} \\
\includegraphics[width=3.25in]{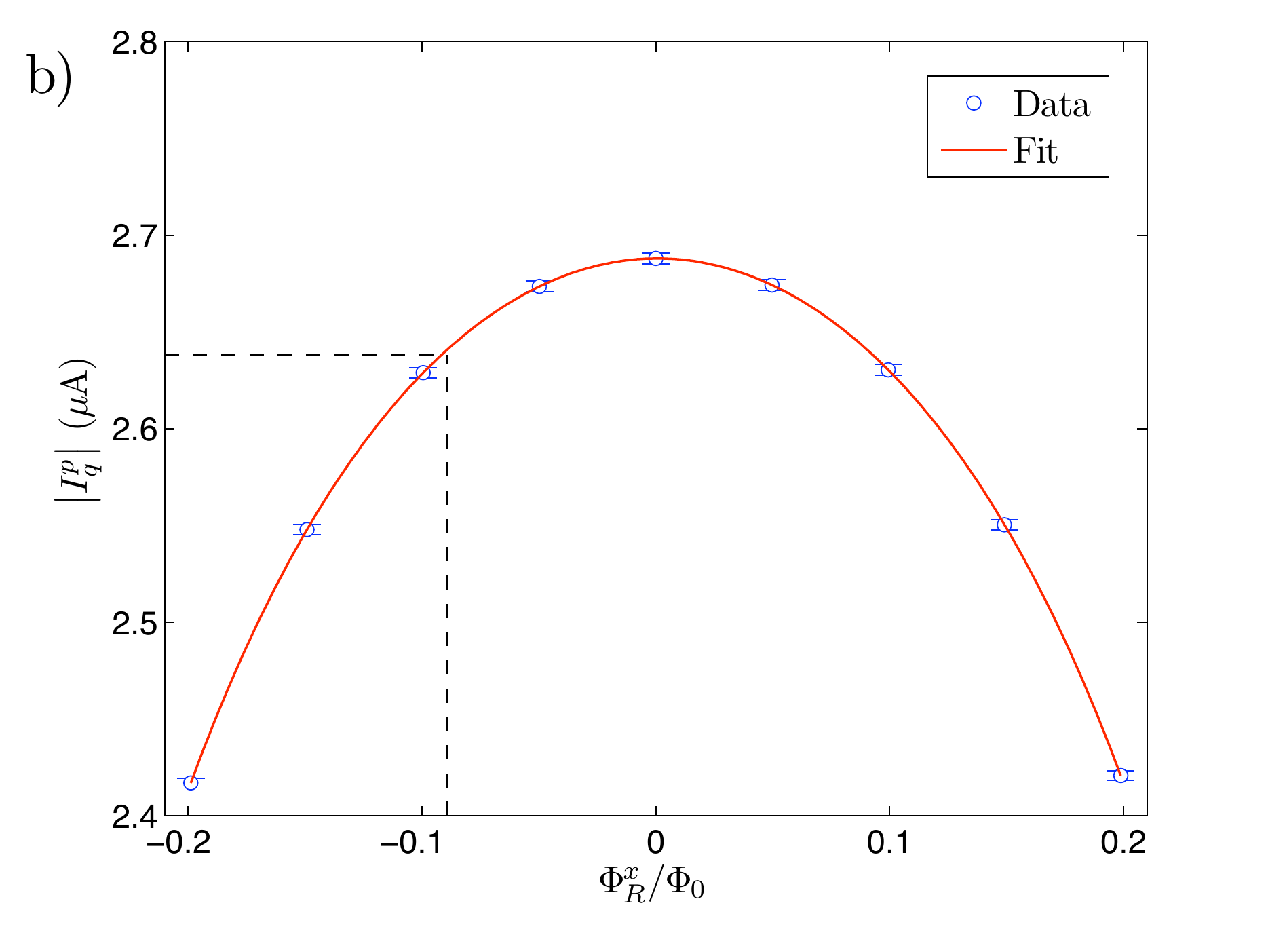} \\
\caption{\label{fig:balanced} (color online) a) Minor lobe balancing data and fit to Eq.~(\ref{eqn:balancedapprox}).  The standard bias conditions for $\Philx$ and $\Phirx$ are indicated by dashed lines. b) $\left|I_q^p(\Phiccjjx=-\Phi_0)\right|$ versus $\Phirx$, where $\Philx$ has been chosen using Eq.~(\ref{eqn:balancedapprox}).  The data have been fit to the ideal CCJJ rf-SQUID model.  The standard bias condition for $\Phirx$ and the resultant $\left|I_q^p(\Phiccjjx=-\Phi_0)\right|$ are indicated by dashed lines.}
\end{figure}

To find balanced pairs of  $\left(\Philx,\Phirx\right)$ in practice, we set $\Phirx$ to a constant and used the LF flux noise rejecting procedure inside a software feedback loop that controlled $\Philx$ to null the difference in apparent degeneracy point to a precision of $20\,\mu\Phi_0$.  Balanced pairs of $\left(\Philx,\Phirx\right)$ are plotted in Fig.~\ref{fig:balanced}a.  These data have been fit to \ref{eqn:balancedapprox} using $\beta_-/\beta_+$ as a free parameter.  The best fit shown in Fig.~\ref{fig:balanced}a was obtained with $1-\beta_{R,+}/\beta_{L,+}=(4.1\pm0.3)\times 10^{-3}$, which then indicates an approximately $0.4\%$ asymmetry between the pairs of junctions in the $L$ and $R$ loops.

A demonstration of how the CCJJ facilitates tuning of $I_q^c$ is shown in Fig.~\ref{fig:balanced}b.  Here, the measurable consequence of altering $I_q^c$ that was recorded was a change in $\iqp$ at $\Phiccjjx=-\Phi_0$.  These data have been fit to the ideal CCJJ rf-SQUID model with the substitution
\begin{equation}
\label{eqn:Icbalanced}
I_q^c(\Phirx,\Philx)=I_c^0\cos\left(\frac{\pi\Phirx}{\Phi_0}\right)
\end{equation}

\noindent and using the values of $L_{\text{ccjj}}$ and $L_q$ obtained from fitting the data in Fig.~\ref{fig:Lq0extraction}, but treating $I_c^0$ as a free parameter.  Here, $\Philx$ on the left side of Eq.~(\ref{eqn:Icbalanced}) is a function of $\Phirx$ per the CCJJ balancing condition Eq.~(\ref{eqn:balancedapprox}).  The best fit was obtained with $I_c^0=3.25\pm0.01\,\mu$A.  This latter quantity agrees well with the designed critical current of four $0.6\,\mu$m diameter junctions in parallel of $3.56\;\mu$A.  Thus, it is possible to target a desired $I_q^c$ by using Eq.~(\ref{eqn:Icbalanced}) to select $\Phirx$ and then Eq.~(\ref{eqn:balancedapprox}) to select $\Philx$.  The standard bias conditions for $\Philx$ and $\Phirx$ quoted previously were chosen so as to homogenize $I_q^c$ amongst the 8 CCJJ rf-SQUIDs on this particular chip.

\subsection{$L$-Tuner}

\begin{figure}[ht]
\includegraphics[width=3.25in]{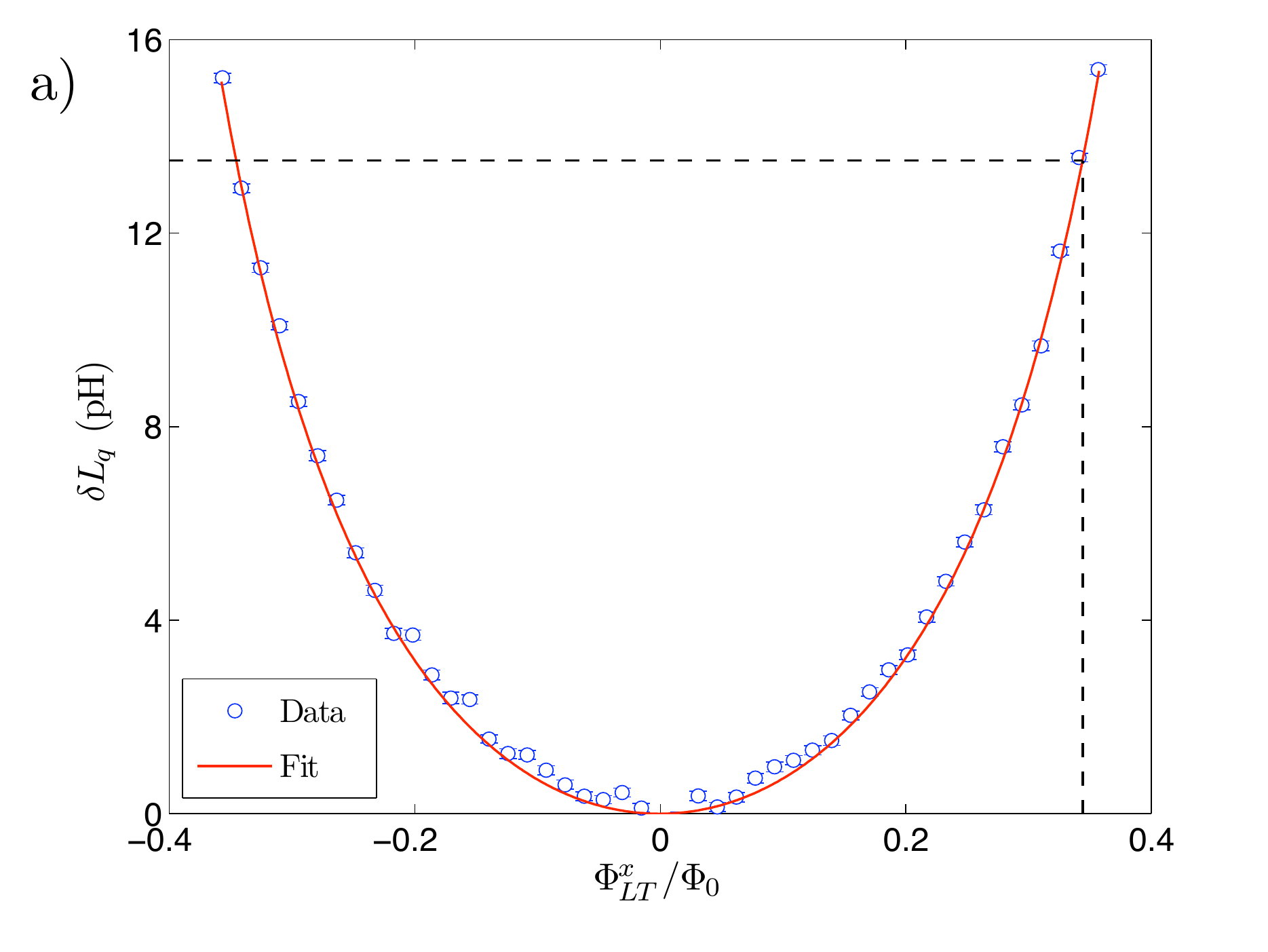} \\
\includegraphics[width=3.25in]{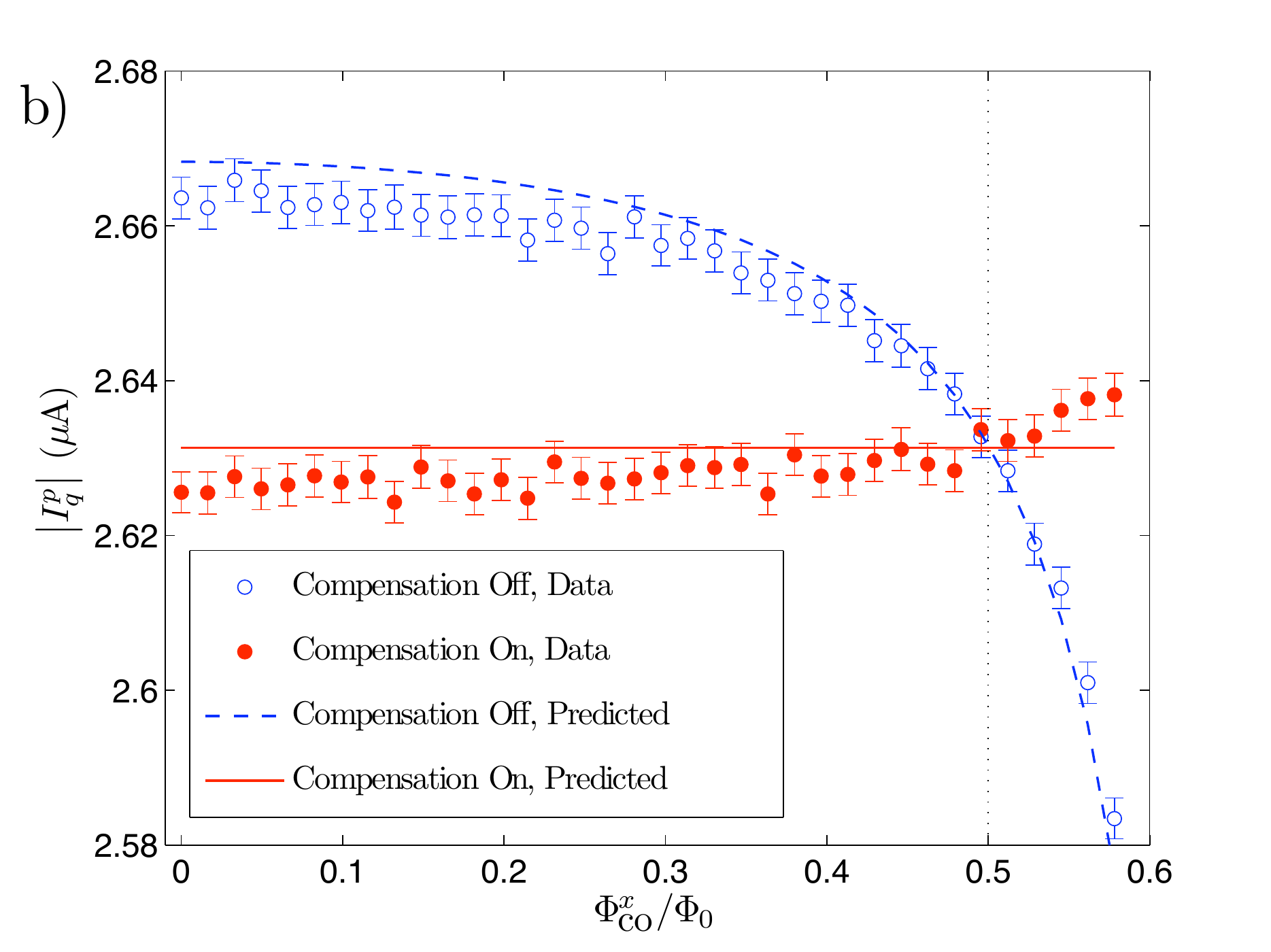} \\
\caption{\label{fig:ltuner} (color online) a) $L$-tuner calibration and fit to Eq.~(\ref{eqn:Ltunerfit}). The standard bias condition for $\Phi^x_{\text{LT}}$ and the resultant $\delta L_q$ are indicated by dashed lines.  b)  Observed change in maximum qubit persistent current with and without active $L$-tuner compensation and predictions for both cases.}
\end{figure}

To characterize the $L$-tuner, we once again turned to measurements of $\left|I_q^p(\Phiccjjx=-\Phi_0)\right|$, but this time as a function of $\Phi^x_{LT}$.  Persistent current results were then used to infer $\delta L_q=L_q(\Phi^x_{LT})-L_q(\Phi^x_{LT}=0)$ using the ideal CCJJ rf-SQUID model with $L_{\text{ccjj}}$ and $I_q^c$ held constant and treating $L_q$ as a free parameter.  The experimental results are plotted in Fig.~\ref{fig:ltuner}a and have been fit to
\begin{equation}
\label{eqn:Ltunerfit}
\delta L_q=\frac{L_{J0}}{\cos\left(\pi\Phi^x_{LT}/\Phi_0\right)} \; ,
\end{equation}

\noindent and the best fit was obtained with $L_{J0}=19.60\pm0.04\,$pH.  Modeling this latter parameter as $L_{q0}=\Phi_0/2\pi I^c_{LT}$, we estimate $I^c_{LT}=16.79\pm0.04\,\mu$A, which is close to the design value of $16.94\,\mu$A.  The standard bias condition for $\Phi^x_{\text{LT}}$ was chosen so as to homogenize $L_q$ amongst the 8 CCJJ rf-SQUID flux qubits on this chip and to provide adequate bipolar range to accommodate inter-qubit coupler operation.

To demonstrate the use of the $L$-tuner, we have probed a worst-case scenario in which four CJJ rf-SQUID couplers connected to the CCJJ rf-SQUID in question are tuned in unison.  Each of the couplers had been independently calibrated per the procedures described in Ref.~\onlinecite{cjjcoupler}, from which we obtained $M_{\text{co},i}\approx 15.8\,$pH and $\chi_i\left(\Phicox\right)$ ($i\in\left\{1,2,3,4\right\}$).  Each of these devices provided a maximum AFM inter-qubit mutual inductance $M_{\text{AFM}}=M^2_{\text{co},i}\chi_{\text{AFM}}\approx 1.56\,$pH, from which one can estimate $\chi_{\text{AFM}}\approx 6.3\,$nH$^{-1}$.  Measurements of $\iqp$ with and without active $L$-tuner compensation as a function of coupler bias $\Phicox$, as applied to all four couplers simultaneously, are presented in Fig.~\ref{fig:ltuner}b.  The predictions from the ideal CCJJ rf-SQUID model, obtained by using $L_q=265.4\,\text{pH}$ (with compensation) and $L_q$ obtained from Eq.~(\ref{eqn:LqNoTuner}) (without compensation), are also shown.  Note that the two data sets and predictions all agree to within experimental error at $\Phicox=0.5\,\Phi_0$, which corresponds to the all zero coupling state ($M_{\text{eff}}=0$).  The experimental results obtained without $L$-tuner compensation agree reasonably well with the predicted $\Phicox$-dependence.  As compared to the case without compensation, it can be seen that the measured $\iqp$ show considerably less $\Phicox$-dependence when $L$-tuner compensation is provided.  However, the data suggest a small systematic deviation from the inductance models Eqs.~(\ref{eqn:LqNoTuner}) and (\ref{eqn:LqWithTuner}).  At $\Phiccjjx=-\Phi_0$, for which it is estimated that $\beta_{\text{eff}}\approx 2.43$, $\iqp\propto 1/L_q$.  Given that the data for the case without compensation are below the model, then it appears that we have slightly underestimated the change in $L_q$.  Consequently, we have provided insufficient ballast inductance when the $L$-tuner compensation was activated.

\subsection{rf-SQUID Capacitance}

Since $I_q^c$ and $L_q$ directly impact the CCJJ rf-SQUID potential in Hamiltonian (\ref{eqn:4JHeff}), it was possible to infer CCJJ and $L$-tuner properties from measurements of the groundstate persistent current.  In contrast, the rf-SQUID capacitance $C_q$ appears in the kinetic term in Hamiltonian (\ref{eqn:4JHeff}).  Consequently, one must turn to alternate experimental methods that invoke excitations of the CCJJ rf-SQUID in order to characterize $C_q$.  One such method is to probe macroscopic resonant tunneling (MRT) between the lowest lying state in one well into either the lowest order [LO, $n=0$] state or into a higher order [HO, $n>0$] state in the opposing well of the rf-SQUID double well potential \cite{HOMRT}.  The spacing of successive HOMRT peaks as a function of rf-SQUID flux bias $\Phiqx$ will be particularly sensitive to $C_q$.  HOMRT has been observed in many different rf-SQUIDs and is a well established quantum mechanical phenomenon \cite{HOMRT,Bennett,MRT3JJ}.  LOMRT proved to be more difficult to observe in practice and was only reported upon relatively recently in the literature \cite{LOMRT}.  We refer the reader to this latter reference for the experimental method for measuring MRT rates.

\begin{figure}
\includegraphics[width=3.25in]{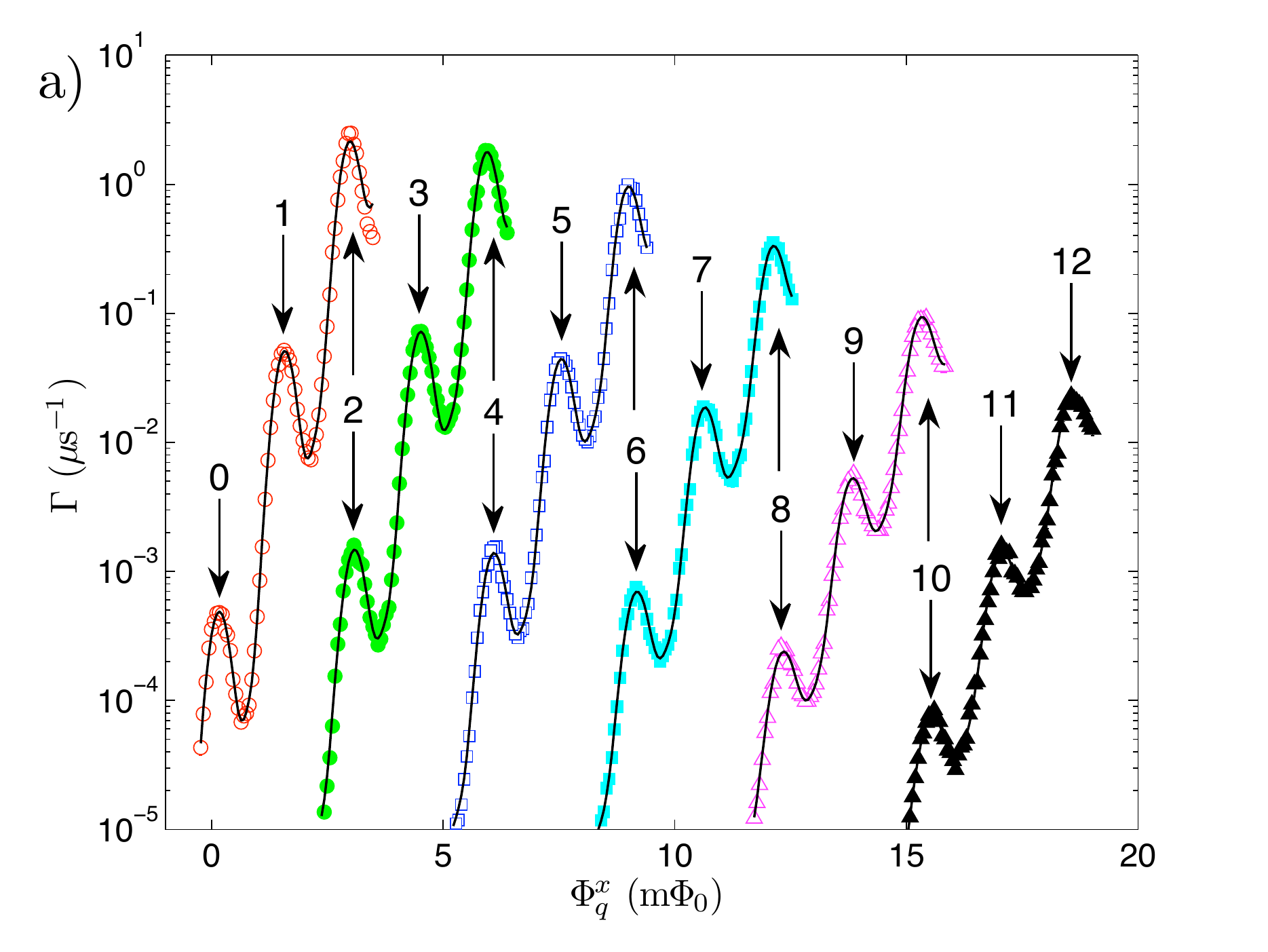} \\
\includegraphics[width=3.25in]{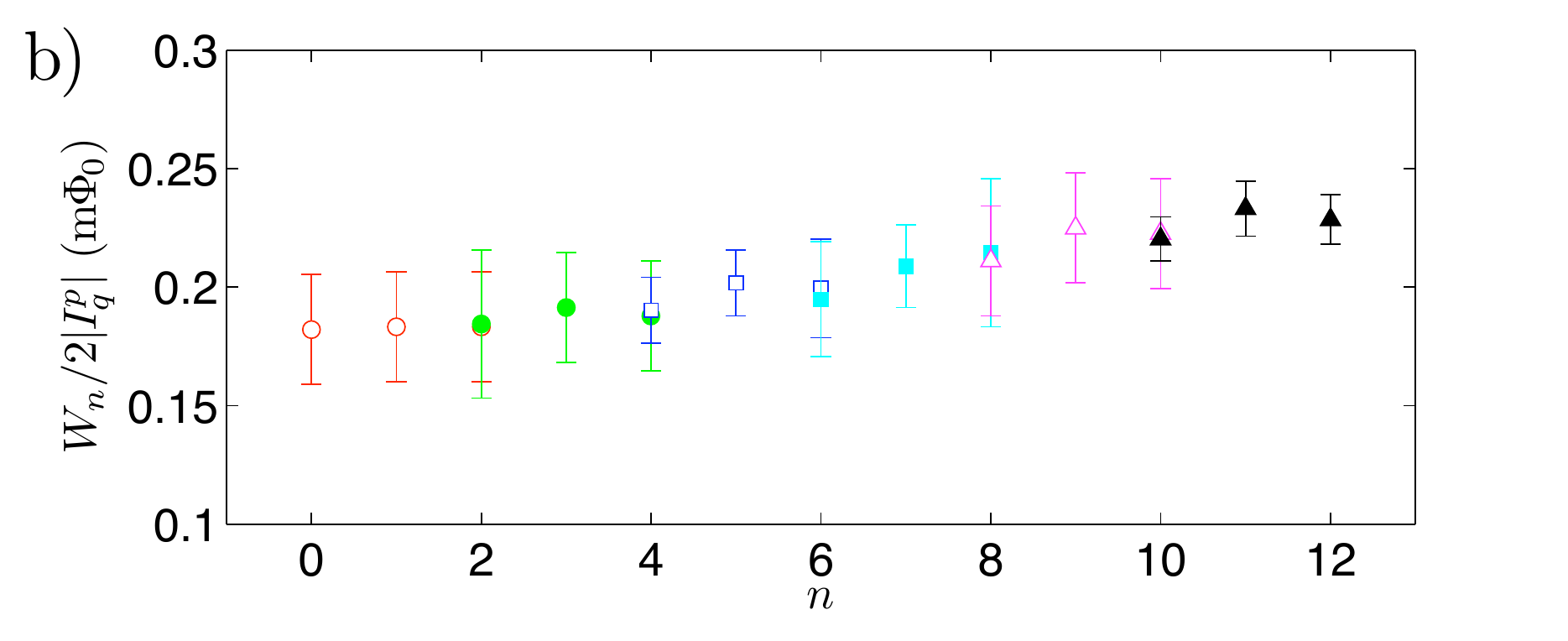} \\
\includegraphics[width=3.25in]{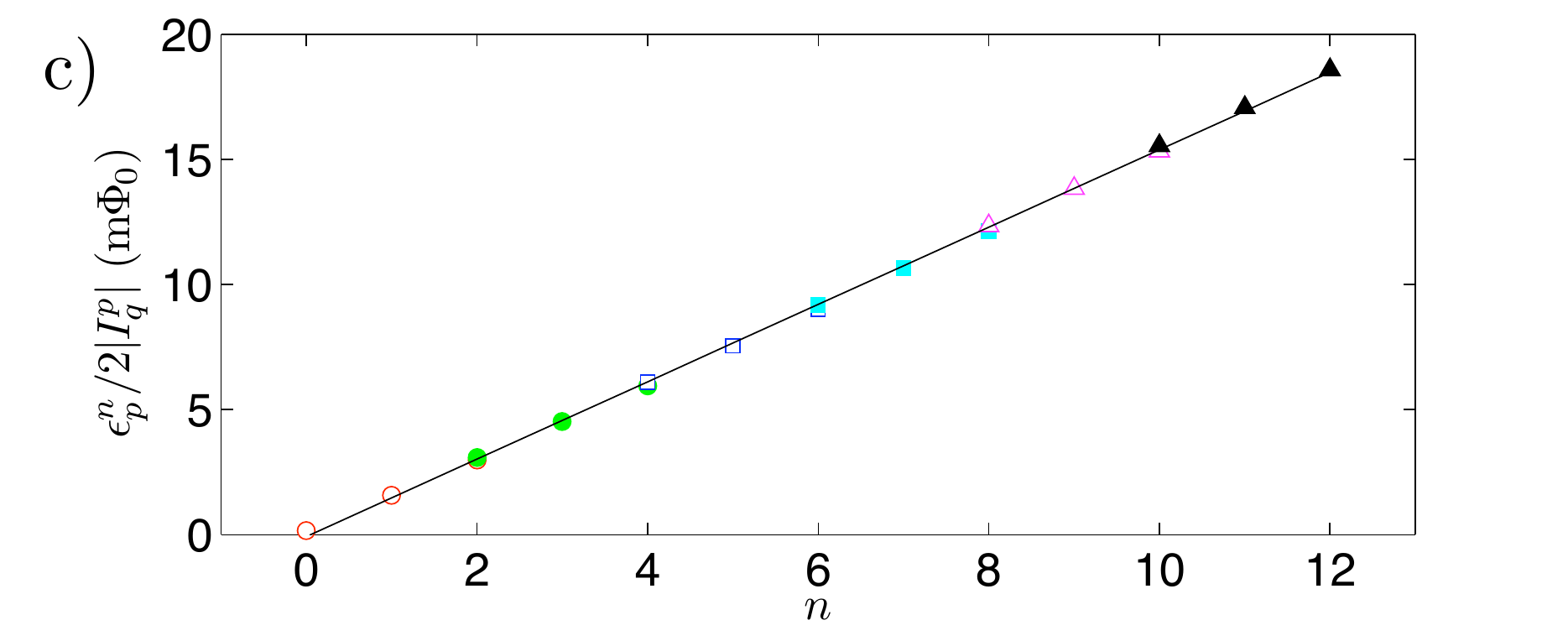}
\caption{\label{fig:HOMRT} (color online) a) HOMRT peaks fitted to Eq.~(\ref{eqn:HOMRTFit}).  Data shown are for $\Phiccjjx/\Phi_0=-0.6677$, $-0.6735$, $-0.6793$, $-0.6851$, $-0.6911$ and $-0.6970$, from left to right, respectively.  Number of levels in target well $n$ as indicated.  b)  Best fit Gaussian width parameter $W_n$ as a function of $n$.  c)  Best fit peak position $\epsilon_p^n$ as a function of $n$.}
\end{figure}

Measurements of the initial decay rate $\Gamma\equiv dP_{\downarrow}/dt|_{t=0}$ versus $\Phiqx$ are shown in Fig.~\ref{fig:HOMRT}a with the order of the target level $n$ as indicated.  The maximum observable $\Gamma$ was imposed by the bandwidth of the apparatus, which was $\sim 5\,$MHz.  The minimum observable $\Gamma$ was dictated by experimental run time constraints.  In order to observe many HO resonant peaks within our experimental bandwidth we have successively raised the tunnel barrier height in roughly equal intervals by tuning the target $\Phiccjjx$.  The result is a cascade of resonant peaks atop a monotonic background.

The authors of Ref.~\onlinecite{Bennett} attempted to fit their HOMRT data to a sum of gaussian broadened lorentzian peaks.  It was found that they could obtain satisfactory fits within the vicinity of the tops of the resonant features but that the model was unable to correctly describe the valleys between peaks.  We had reached the same conclusion with the very same model as applied to our data. However, it was empirically observed that we could obtain excellent fits to all of the data by using a model composed of a sum of purely gaussian peaks plus a background that varies exponentially with $\Phiqx$:
\begin{equation}
\label{eqn:HOMRTFit}
\frac{\Gamma(\Phiqx)}{\hbar}=\sum_{n}\sqrt{\frac{\pi}{8}}\frac{\Delta_n^2}{W_n}e^{-\frac{(\epsilon-\epsilon_p^n)^2}{2W_n^2}}+\Gamma_{\text{bkgd}}e^{\Phiqx/\delta\Phi_{\text{bkgd}}}\; ,
\end{equation}

\noindent where $\epsilon\equiv 2\iqp\Phiqx$. These fits are shown in Fig.~\ref{fig:HOMRT}a.  A summary of the gaussian width parameter $W_n$ in Fig.~\ref{fig:HOMRT}b is shown solely for informational purposes. We will refrain from speculating why there is no trace of lorentzian lineshapes or on the origins of the exponential background herein, but rather defer a detailed examination of HOMRT to a future publication.

For the purposes of this article, the key results to take from the fits shown in Fig.~\ref{fig:HOMRT}a are the positions of the resonant peaks, as plotted in Fig.~\ref{fig:HOMRT}c.  These results indicate that the peak spacing is very uniform: $\delta\Phi_{\text{MRT}}=1.55\pm0.01\,$m$\Phi_0$.  One can compare $\delta\Phi_{\text{MRT}}$ with the predictions of the ideal CCJJ rf-SQUID model using the previously calibrated $L_q=265.4\,$pH, $L_{\text{ccjj}}=26\,$pH and $I_q^c=3.103\,\mu$A  with $C_q$ treated as a free parameter.  From such a comparison, we estimate $C_q=190\pm 2\,$fF.  

The relatively large value of $C_q$ quoted above can be reconciled with the CCJJ rf-SQUID design by noting that, unlike other rf-SQUID flux qubits reported upon in the literature, our qubit body resides proximal to a superconducting groundplane so as to minimize crosstalk.  In this case, the qubit wiring can be viewed as a differential transmission line of length $\ell/2\sim 900\,\mu$m, where $\ell$ is the total length of qubit wiring, with the effective Josephson junction and a short on opposing ends.  The transmission line will present an impedance of the form $Z(\omega)=-j Z_0\tanh(\omega\ell/2\nu)$ to the effective Josephson junction, with the phase velocity $\nu\equiv1/\sqrt{L_0C_0}$ defined by the differential inductance per unit length $L_0\sim 0.26\,$pH$/\mu$m and capacitance per unit length $C_0\sim 0.18\,$fF$/\mu$m, as estimated from design.  If the separation between differential leads is greater than the distance to the groundplane, then $\ell/2\nu\approx\sqrt{L_{\text{body}}C_{\text{body}}/4}$, where $C_{\text{body}}\sim 640\,$fF is the total capacitance of the qubit wiring to ground.  Thus, one can model the high frequency behavior of the shorted differential transmission line as an inductance $L_{\text{body}}$ and a capacitance $C_{\text{body}}/4$ connected in parallel with the CCJJ.  Taking a reasonable estimated value of $40\,$fF/$\mu$m$^2$ for the capacitance per unit area of a Josephson junction, one can estimate the total capacitance of four $0.6\,\mu$m diameter junctions in parallel to be $C_J\sim 45\,$fF.  Thus we estimate $C_q=C_J+C_{\text{body}}/4\sim 205\,$fF, which is in reasonable agreement with the best fit value of $C_q$ quoted above.

With all of the controls of the CCJJ rf-SQUID having been demonstrated, we reach the first key conclusion of this article: The CCJJ rf-SQUID is a robust device in that parametric variations, both within an individual device and between a multitude of such devices, can be accounted for using purely static flux biases.  These biases have been applied to all 8 CCJJ rf-SQUIDs on this particular chip using a truly scalable architecture involving on-chip flux sources that are programmed by only a small number of address lines \cite{PCC}.

\section{Qubit Properties}

The purpose of the CCJJ rf-SQUID is to provide an as ideal as possible flux qubit \cite{fluxqubit}.  By this statement, it is meant that the physics of the two lowest lying states of the device can be described by an effective Hamiltonian of the form Eq.~(\ref{eqn:Hqubit}) with $\epsilon=2\iqp\left(\Phi_q^x-\Phi^0_q\right)$,  $\iqp$ being the magnitude of the persistent current that flows about the inductive loop when the device is biased hard to one side, $\Phi_q^0$ being a static flux offset and $\Delta_q$ representing the tunneling energy between the lowest lying states when biased at its degeneracy point $\Phi_q^x=\Phi^0_q$.  Thus, $\iqp$ and $\Delta_q$ are the defining properties of a flux qubit, regardless of its topology \cite{Leggett}.  Given the complexity of a six junction device with five closed superconducting loops, it is quite justifiable to question whether the CCJJ rf-SQUID constitutes a qubit.  These concerns will be directly addressed herein by demonstrating that measured $\iqp$ and $\Delta_q$ agree with the predictions of the quantum mechanical Hamiltonian (\ref{eqn:4JHeff}) given independently calibrated values of $L_q$, $L_{\text{ccjj}}$, $I_q^c$ and $C_q$.

Before proceeding, it is worth providing some context in regards to the choice of experimental methods that have been described below.  For those researchers attempting to implement GMQC using resonant electromagnetic fields to prepare states and mediate interactions between qubits, experiments that involve high frequency pulse sequences to drive excitations in the qubit (such as Rabi oscillations\cite{MooijSuperposition}, Ramsey fringes\cite{MooijSuperposition,1OverFFluxQubit1} and spin-echo\cite{MooijSuperposition,1OverFFluxQubit1,1OverFFluxQubit2}) are the natural modality for studying quantum effects.  Such experiments are convenient in this case as the methods can be viewed as basic gate operations within this intended mode of operation.  However, such methods are not the exclusive means of characterizing quantum resources.  For those who wish to use precise dc pulses to implement GMQC or whose interests lie in developing hardware for AQO, it is far more convenient to have a set of tools for characterizing quantum mechanical properties that require only low bandwidth bias controls.  Such methods, some appropriate in the coherent regime \cite{Greenberg,gsip} and others in the incoherent regime \cite{HOMRT,LOMRT,LZ}, have been reported in the literature.  We have made use of such low frequency methods as our apparatuses typically possess 128 low bandwidth bias lines to facilitate the adiabatic manipulation of a large number of devices.

\begin{figure}
\includegraphics[width=3.25in]{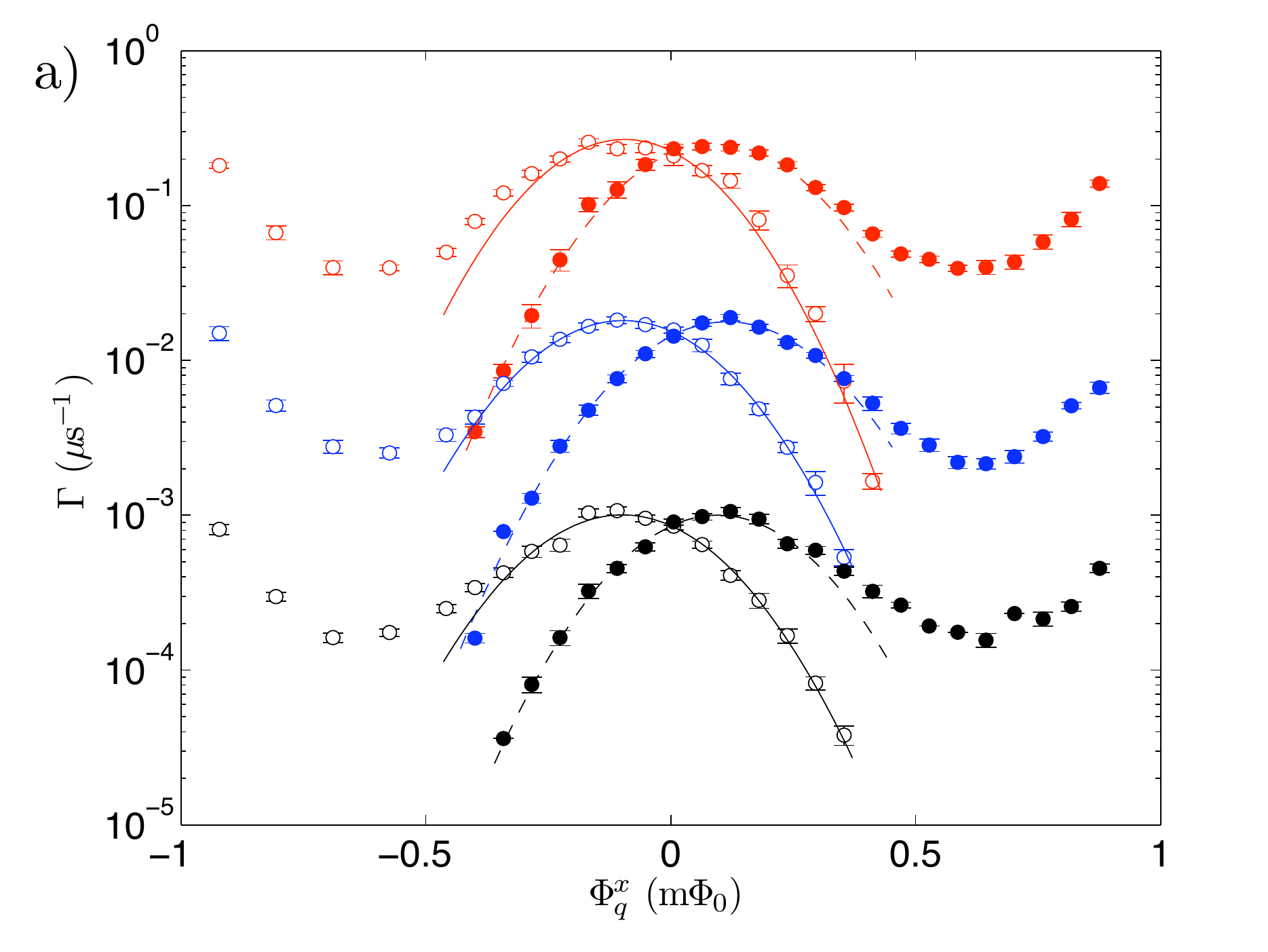} \\
\includegraphics[width=3.25in]{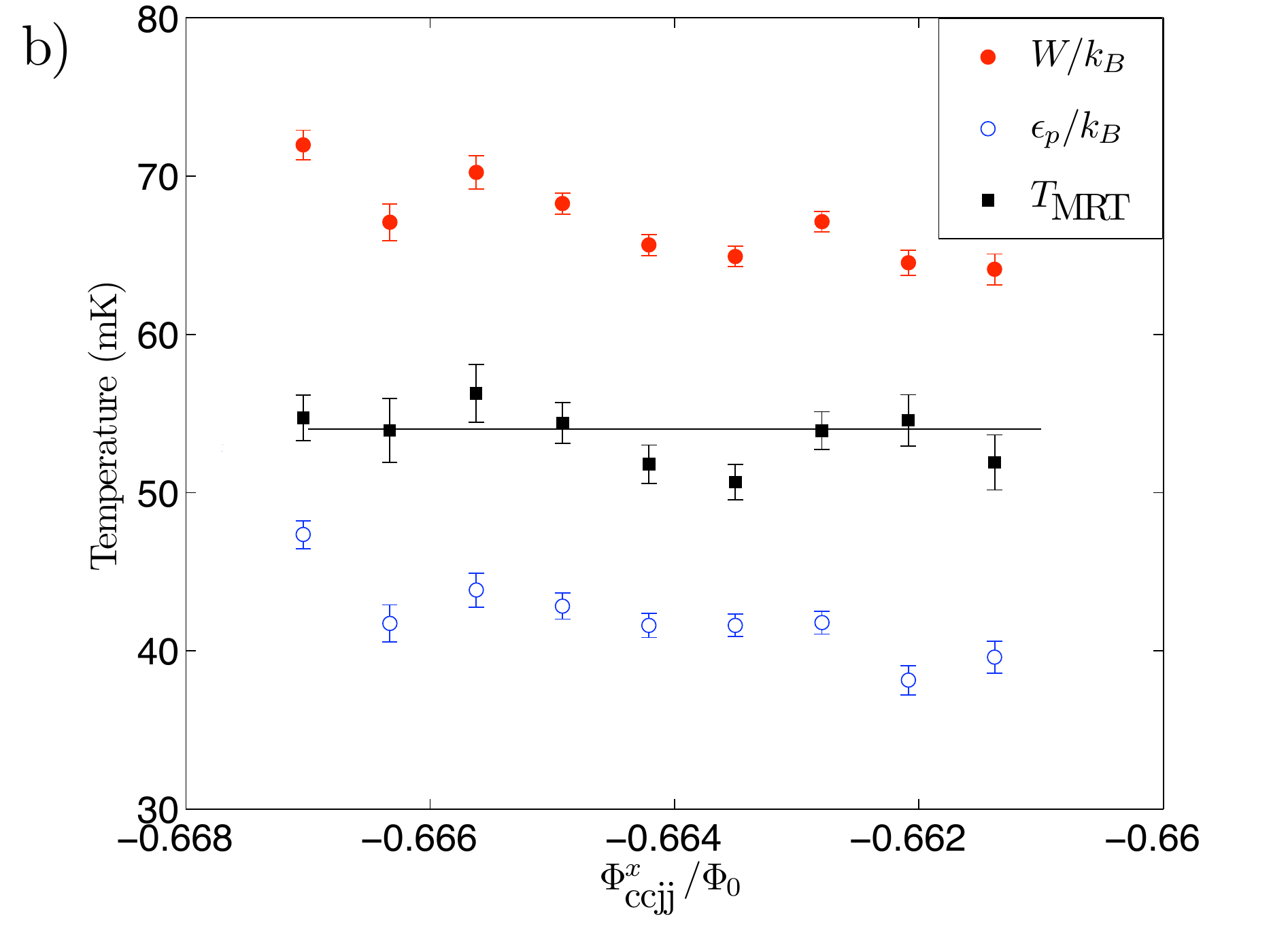} \\
\caption{\label{fig:LOMRT} (color online) a) Example LOMRT peaks fitted to Eq.~(\ref{eqn:LOMRTFit}).  Data shown are for $\Phiccjjx/\Phi_0=-0.6621$, $-0.6642$ and $-0.6663$, from top to bottom, respectively.  Data from the qubit initialized in $\ket{\downarrow}$ ($\ket{\uparrow}$) are indicated by solid (hollow) points.  b) Energy scales obtained from fitting multiple LOMRT traces.}
\end{figure}

One possible means of probing quantum mechanical tunneling between the two lowest lying states of a CCJJ rf-SQUID is via MRT\cite{LOMRT}.  Example LOMRT decay rate data are shown in Fig.~\ref{fig:LOMRT}a.  We show results for both initializations, $\ket{\downarrow}$ and $\ket{\uparrow}$, and fits to gaussian peaks, as detailed in Ref.~\onlinecite{LOMRT}:
\begin{equation}
\label{eqn:LOMRTFit}
\frac{\Gamma(\Phiqx)}{\hbar}=\sqrt{\frac{\pi}{8}}\frac{\Delta_q^2}{W}e^{-\frac{(\epsilon-\epsilon_p)^2}{2W^2}} \;\; .
\end{equation}

\noindent A summary of the fit parameters $\epsilon_p$ and $W$ versus $\Phiccjjx$ is shown in Fig.~\ref{fig:LOMRT}b.  We also provide estimates of the device temperature using the formula
\begin{equation}
\label{eqn:TMRT}
k_BT_{\text{MRT}}=\frac{W^2}{2\epsilon_p} \; .
\end{equation}

\noindent  As expected, $T_{\text{MRT}}$ shows no discernible $\Phiccjjx$-dependence and is scattered about a mean value of $53\pm2\,$mK.  A summary of $\Delta_q$ versus $\Phiccjjx$ will be shown in conjunction with more experimental results at the end of this section.  For further details concerning LOMRT, the reader is directed to Ref.~\onlinecite{LOMRT}.

A second possible means of probing $\Delta_q$ is via a Landau-Zener experiment \cite{LZ}.  In principle, this method should be applicable in both the coherent and incoherent regime.  In practice, we have found it only possible to probe the device to modestly larger $\Delta_q$ than we can reach via LOMRT purely due to the low bandwidth of our bias lines.  Results from such experiments on the CCJJ rf-SQUID flux qubit will be summarized at the end of this section.  We see no fundamental limitation that would prevent others with higher bandwidth apparatuses to explore the physics of the CJJ or CCJJ flux qubit at the crossover between the coherent and incoherent regimes using the Landau-Zener method.  

In order to probe the qubit tunnel splitting in the coherent regime using low bandwidth bias lines, we have developed a new experimental procedure for sensing the expectation value of the qubit persistent current, similar in spirit to other techniques already reported in the literature \cite{gsip}.  An unfortunate consequence of the choice of design parameters for our high fidelity QFP-enabled readout scheme is that the QFP is relatively strongly coupled to the qubit, thus limiting its utility as a detector when the qubit tunnel barrier is suppressed.  One can circumvent this problem within our device architecture by tuning an inter-qubit coupler to a finite inductance and using a second qubit as a latching sensor, in much the same manner as a QFP.  Consider two flux qubits coupled via a mutual inductance $M_{\text{eff}}$.  The system Hamiltonian can then be modeled as
\begin{equation}
\label{eqn:H2Q}
{\cal H}=-\sum_{i\in\left\{q,d\right\}}\frac{1}{2}\left[\epsilon_i\sigma_z^{(i)}+\Delta_i\sigma_x^{(i)}\right]+J\sigma_z^{(q)}\sigma_z^{(d)} \; ,
\end{equation}

\noindent where $J\equiv M_{\text{eff}}|I_q^p||I_d^p|$.  Let qubit $q$ be the flux source and qubit $d$ serve the role of the detector whose tunnel barrier is adiabatically raised during the course of a measurement, just as in a QFP single shot measurement depicted in Fig.~\ref{fig:unipolarannealing}.  In the limit $\Delta_d\rightarrow 0$ one can write analytic expressions for the dispersion of the four lowest energies of Hamiltonian (\ref{eqn:H2Q}):
\begin{equation}
\label{eqn:E2Q}
\begin{array}{ccc}
E_{1\pm} & =  & \pm\frac{1}{2}\sqrt{\left(\epsilon_q-2J\right)^2+\Delta_1^2}-\frac{1}{2}\epsilon_d \; ;\\
E_{2\pm} & =  & \pm\frac{1}{2}\sqrt{\left(\epsilon_q+2J\right)^2+\Delta_1^2}+\frac{1}{2}\epsilon_d \; .
\end{array}
\end{equation}

\noindent As with the QFP, let the flux bias of the detector qubit be engaged in a feedback loop to track its degeneracy point where $P_{d,\downarrow}=1/2$.  Assuming Boltzmann statistics for the thermal occupation of the four levels given by Eq.~(\ref{eqn:E2Q}), this condition is met when
\begin{equation}
\label{eqn:P2minus}
P_{d,\downarrow}=\frac{1}{2}=\frac{e^{-E_{2-}/k_BT}+e^{-E_{2+}/k_BT}}{\sum_{\alpha\in\left\{1\pm,2\pm\right\}}e^{-E_{\alpha}/k_BT}} \; .
\end{equation}

\noindent Setting $P_{d,\downarrow}=1/2$ in Eq.~(\ref{eqn:P2minus}) and solving for $\epsilon_2$ then yields an analytic formula for the balancing condition:
\begin{equation}
\label{eqn:HalfCondGeneral}
\epsilon_d= \frac{F(+)-F(-)}{2}+k_BT\ln\left(\frac{1+e^{-F(+)/k_BT}}{1+e^{-F(-)/k_BT}}\right) \; ;
\end{equation}
\vspace{-0.12in}
\begin{displaymath}
F(\pm)\equiv\sqrt{\left(\epsilon_q\pm 2J\right)^2+\Delta_1^2} \; .
\end{displaymath}

While Eq.~(\ref{eqn:HalfCondGeneral}) may look unfamiliar, it readily reduces to an intuitive result in the limit of small coupling $J\ll \Delta_1$ and $T\rightarrow 0$:
\begin{equation}
\label{eqn:HalfCondSmallJ}
\epsilon_d  \approx  M_{\text{eff}}|I_q^p|\frac{\epsilon_q}{\sqrt{\epsilon_q^2+\Delta_q^2}} = M_{\text{eff}}\bra{g}\hat{I}_q^p\ket{g} \; ,
\end{equation}

\noindent where $\ket{g}$ denotes the groundstate of the source qubit and $\hat{I}_q^p\equiv\iqp\sigma_z^{(q)}$ is the source qubit persistent current operator.  Thus Eq.~(\ref{eqn:HalfCondGeneral}) is an expression for the expectation value of the source qubit's groundstate persistent current in the presence of backaction from the detector and finite temperature.  Setting $\epsilon_i=2|I_i^p|\Phi^x_{i}$ and rearranging then gives an expression for the flux bias of the detector qubit as a function of flux bias applied to the source qubit.  Given independent calibrations of $M_{\text{eff}}=1.56\pm 0.01\,$pH for a particular coupler set to $\Phicox=0$ on this chip, $T=54\pm 3\,$mK from LOMRT fits and $|I_d^p|=1.25\pm0.02\,\mu$A at the CCJJ bias where the LOMRT rate approaches the bandwidth of our bias lines, one can then envision tracing out $\Phi_d^x$ versus $\Phi_q^x$ and fitting to Eq.~(\ref{eqn:HalfCondGeneral}) to extract the source qubit parameters $|I_q^p|$ and $\Delta_q$ .

\begin{figure}
\includegraphics[width=3.25in]{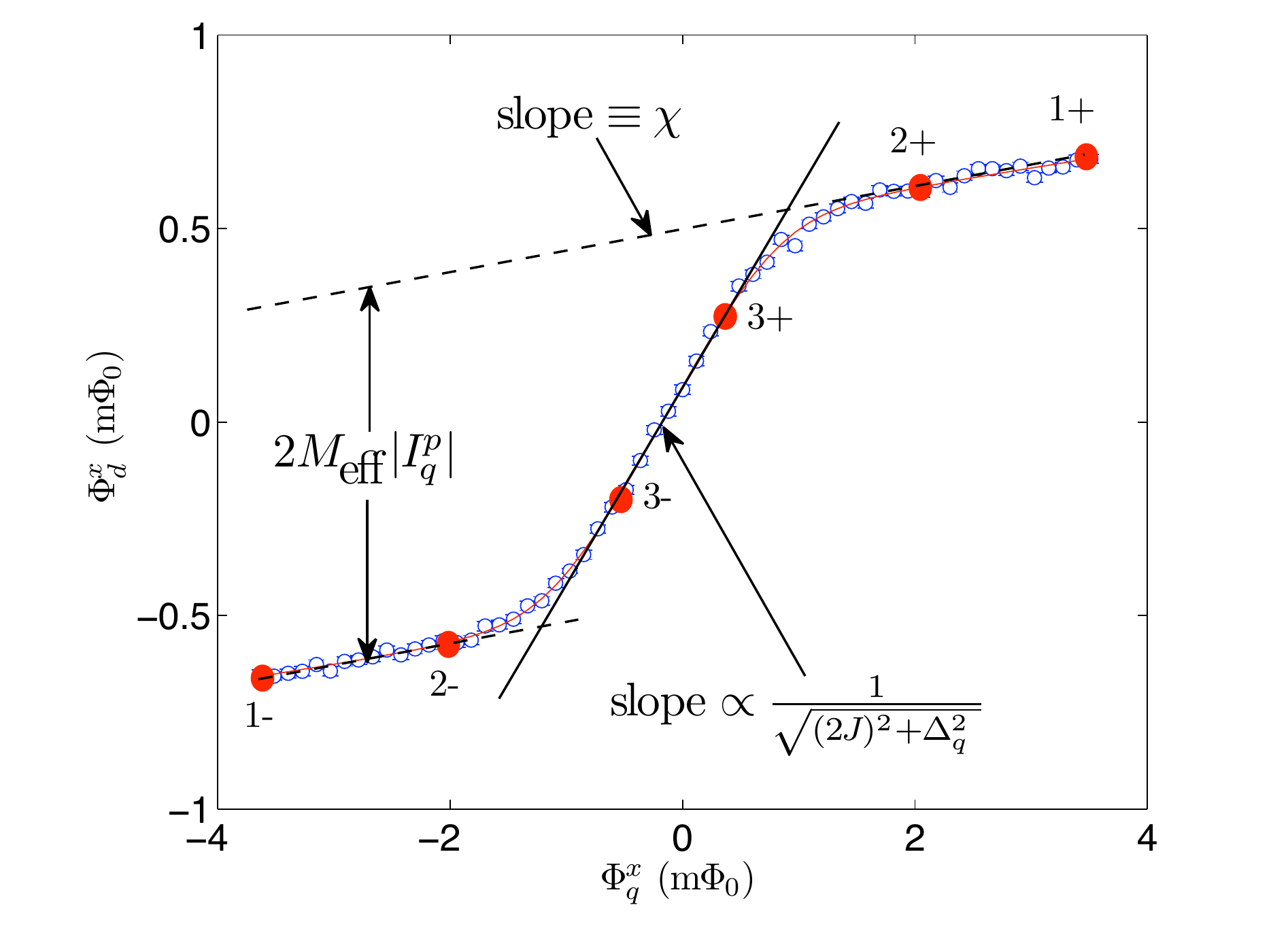}
\caption{\label{fig:largedeltatrace}  (color online)  Example coupled flux trace taken at $\Phiccjjx=-0.6513\,\Phi_0$ used to extract large $\Delta$ in the coherent regime. }
\end{figure}

An example $\Phi_d^x$ versus $\Phi_q^x$ data set for source CCJJ flux bias $\Phiccjjx=-0.6513\,\Phi_0$ is shown in Fig.~\ref{fig:largedeltatrace}.  The solid curve in this plot corresponds to a fit to Eq.~(\ref{eqn:HalfCondGeneral}) with a small background slope that we denote as $\chi$.  We have confirmed from the ideal CCJJ rf-SQUID model that $\chi$ is due to the diamagnetic response of the source rf-SQUID to changing $\Phi_q^x$.  This feature becomes more pronounced with increasing $C_q$ and is peaked at the value of $\Phiccjjx$ for which the source qubit potential becomes monostable, $\beta_{\text{eff}}=1$.  Nonetheless, the model also indicates that $\chi$ in no way modifies the dynamics of the rf-SQUID, thus the qubit model still applies.  From fitting these particular data, we obtained $|I_q^p|=0.72\pm 0.04\,\mu$A and $\Delta_q/h=2.64\pm 0.24\,$GHz.

\begin{figure}
\includegraphics[width=3.25in]{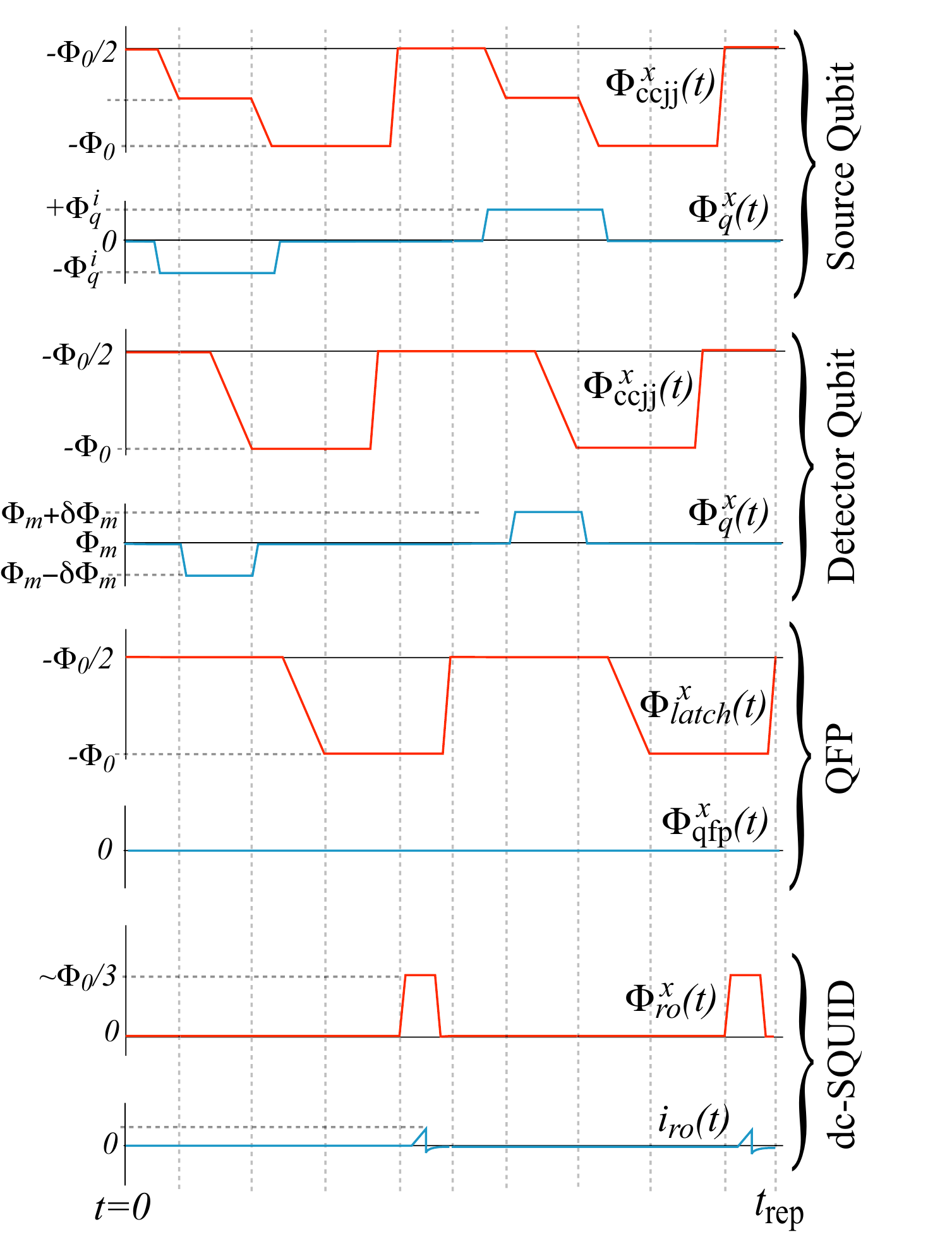}
\caption{\label{fig:deltawaveforms}  (color online) Depiction of large $\Delta_q$ measurement waveforms.  The waveform sequence is similar to that of Fig.~\ref{fig:iqplockin}, albeit the source qubit's tunnel barrier is partially suppressed ($-\Phi_0/2<\Phiccjjx<-\Phi_0$) and  a second qubit (as opposed to a QFP) serves as the flux detector.}
\end{figure}

In practice we have found it inefficient to take detailed traces of $\Phi_d^x$ versus $\Phi_q^x$ as this procedure is susceptible to corruption by LF flux noise in the detector qubit.  As an alternative approach, we have adapted the LF flux noise rejecting procedures introduced in the last section of this article to measure a series of three differential flux levels in the detector qubit.  The waveforms needed to accomplish this task are depicted in Fig.~\ref{fig:deltawaveforms}.  Here, the dc-SQUID and QFP connected to the detector qubit are used in latching readout mode while the detector qubit is annealed in the presence of a differential flux bias $\Phi_m\pm\delta\Phi_m$ which is controlled via feedback.  Meanwhile, the source qubit's CCJJ bias is pulsed to an intermediate level $-\Phi_0<\Phiccjjx<-\Phi_0/2$ in the presence of an initialization flux bias $\pm\Phi_q^i$.  By choosing two appropriate pairs of levels $\pm\Phi_q^i$, as indicated by the solid points $1\pm$ and $2\pm$ in Fig.~\ref{fig:largedeltatrace}, one can extract $\iqp$ and $\chi$ from the two differential flux measurements.  In order to extract $\Delta_q$, we then choose a pair of $\pm\Phi_q^i$ in the centre of the trace, as indicated by the solid points $3\pm$, from which we obtain the central slope $d\Phi_d^x/d\Phi_q^x$.  Taking the first derivative of Eq.~(\ref{eqn:HalfCondGeneral}) and evaluating at $\Phi_q^x=0$ yields
\begin{equation}
\label{eqn:centralslope}
\frac{d\Phi_d^x}{d\Phi_q^x}-\chi=\frac{2M_{\text{eff}}\iqp^2}{\sqrt{\left(2J\right)^2+\Delta_q^2}}\tanh\left[\frac{\sqrt{\left(2J\right)^2+\Delta_q^2}}{2k_bT}\right] \; .
\end{equation}

\noindent Given independent estimates of all other parameters, one can then extract $\Delta_q$ from this final differential flux measurement.

\begin{figure}
\includegraphics[width=3.25in]{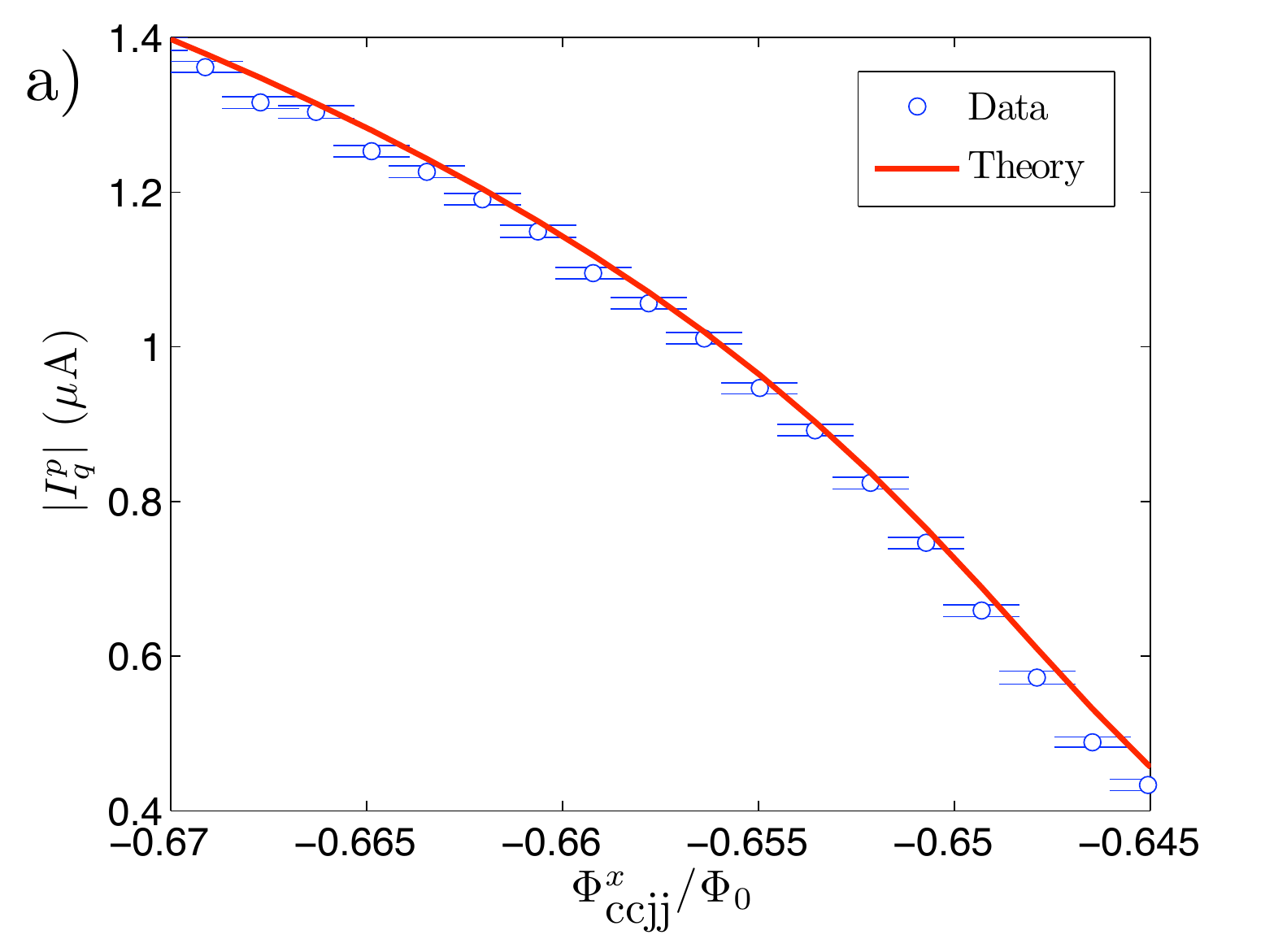} \\
\includegraphics[width=3.25in]{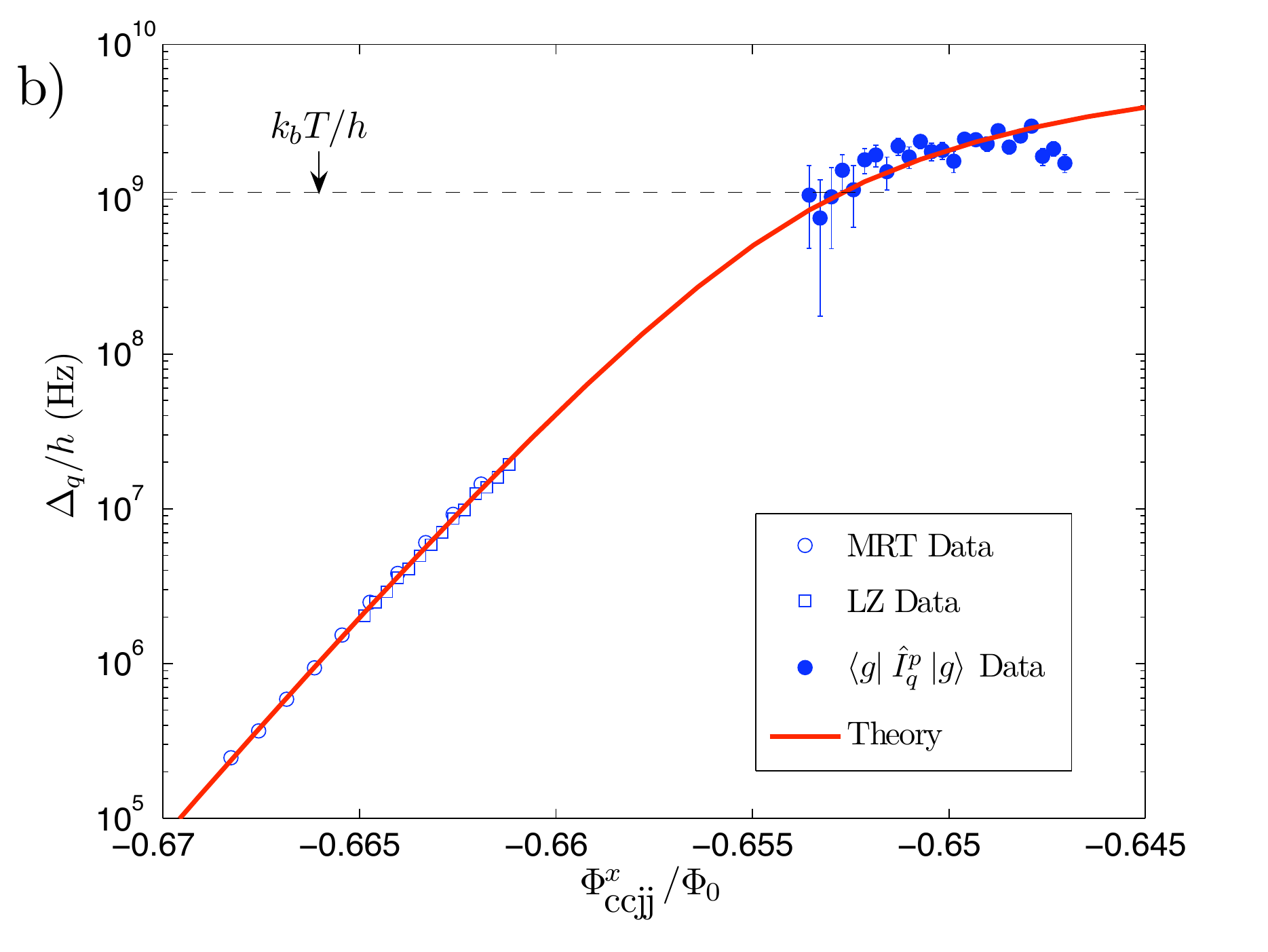}
\caption{\label{fig:DeltaAndIp} (color online) a)  Magnitude of the persistent current $\iqp$ as a function of $\Phiccjjx$.  b)  Tunneling energy $\Delta_q$ between two lowest lying states of the CCJJ rf-SQUID as a function of $\Phiccjjx$, as characterized by macroscopic resonant tunneling [MRT] and Landau-Zener [LZ] in the incoherent regime and coupled groundstate persistent current ($\bra{g}\hat{I}_q^p\ket{g}$) in the coherent regime.  Solid curves are the predictions of the ideal CCJJ rf-SQUID model using independently calibrated $L_q$, $L_{\text{ccjj}}$, $I_q^c$ and $C_q$ with no free parameters.}
\end{figure}

A summary of experimental values of the qubit parameters $\iqp$ and $\Delta_q$ versus $\Phiccjjx$ is shown in Fig.~\ref{fig:DeltaAndIp}.  Here, we have taken $\Delta_q$ from LOMRT and Landau-Zener experiments in the incoherent regime and from the LF flux noise rejecting persistent current procedure discussed above in the coherent regime.  The large gap between the three sets of measurements is due to two reasons: First, the relatively low bandwidth of our bias lines does not allow us to perform MRT or Landau-Zener measurements at higher $\Delta_q$ where the dynamics are faster.  Second, while the coherent regime method worked for $\Delta_q>k_BT$, it proved difficult to reliably extract $\Delta_q$ in the opposite limit.  As such, we cannot make any precise statements regarding the value of $\Phiccjjx$ which serves as the delineation between the coherent and incoherent regimes based upon the data shown in Fig.~\ref{fig:DeltaAndIp}b.  Regulating the device at lower temperature would assist in extending the utility of the coherent regime method to lower $\Delta_q$.  On the other hand, given that Eq.~(\ref{eqn:LOMRTFit}) predicts that $\Gamma\propto\Delta_q^2$, one would have to augment the experimental bandwidth by at least two orders of magnitude to gain one order of magnitude in $\Delta_q$ via either MRT or LZ experiments.

The solid curves in Fig.~\ref{fig:DeltaAndIp} were generated with the ideal CCJJ rf-SQUID model using the independently calibrated $L_q=265.4\,$pH, $L_{\text{ccjj}}=26\,$pH, $I_q^c=3.103\,\mu$A and $C_q=190\,$fF.  Note that there are no free parameters.  It can be seen that the agreement between theory and experiment is quite reasonable.  Thus we reach the second key conclusion of this article: The CCJJ rf-SQUID can be identified as a flux qubit as the measured $\iqp$ and $\Delta_q$ agree with the predictions of a quantum mechanical Hamiltonian whose parameters were independently calibrated.

\section{Noise}

With the identification of the CCJJ rf-SQUID as a flux qubit firmly established, we now turn to assessing the relative quality of this device in comparison to other flux qubits reported upon in the literature.  In this section, we present measurements of the low frequency flux and critical current spectral noise densities, $S_{\Phi}(f)$ and $S_{I}(f)$, respectively.  Finally, we provide explicit links between $S_{\Phi}(f)$ and the free induction (Ramsey) decay time $T^*_{2}$ that would be relevant were this flux qubit to be used as an element in a gate model quantum information processor.

\subsection{Flux Noise}

\begin{figure}
\includegraphics[width=3.25in]{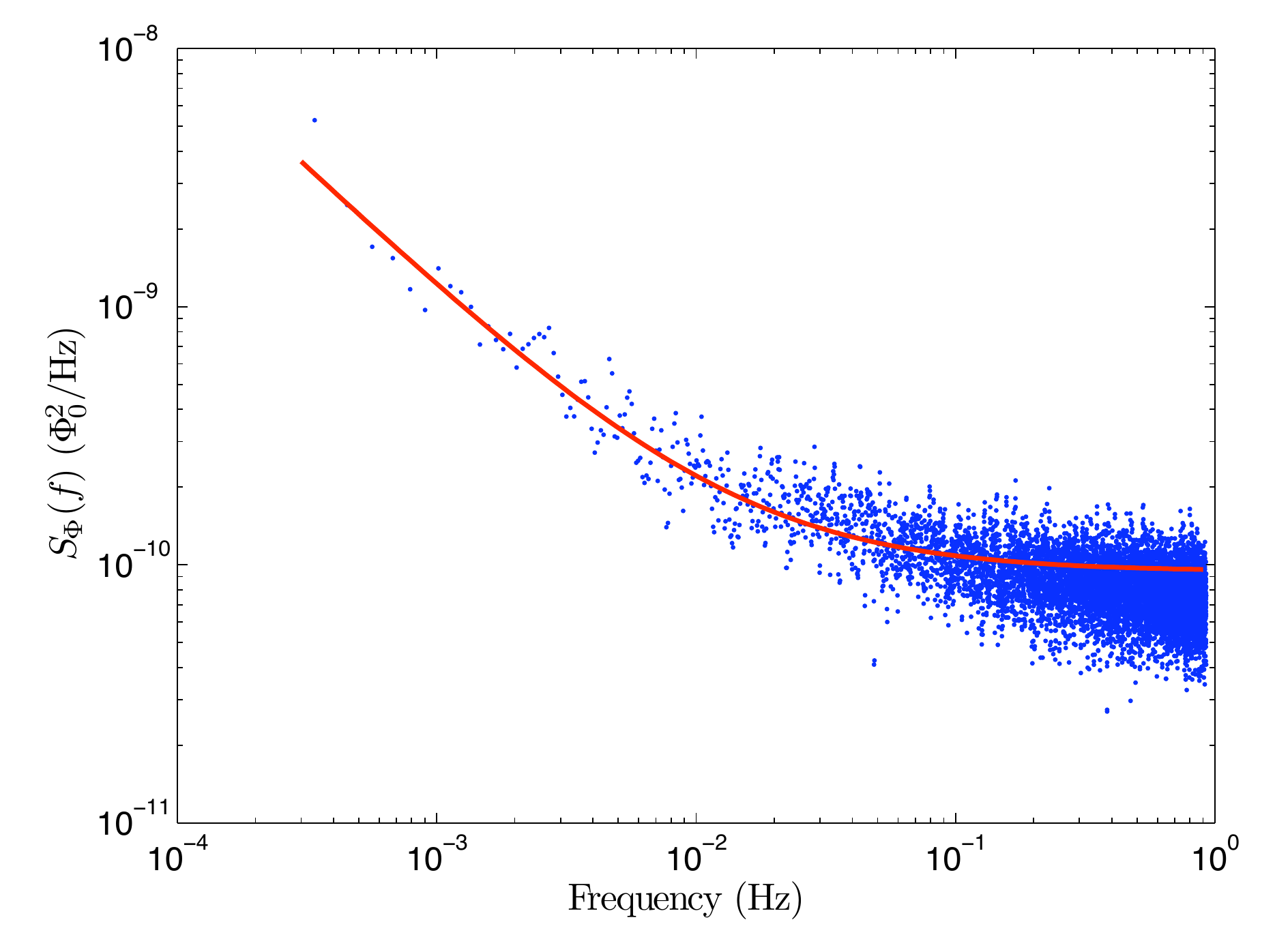}
\caption{\label{fig:fluxnoise} (color online) Low frequency flux noise in the CCJJ rf-SQUID flux qubit.  Data [points] have been fit to Eq.~(\ref{eqn:1OverF}) [solid curve].}
\end{figure}

Low frequency ($1/f$) flux noise is ubiquitous in superconducting devices and is considered a serious impediment to the development of large scale solid state quantum information processors \cite{1OverF}.  We have performed systematic studies of this property using a large number of flux qubits of varying geometry \cite{1OverFGeometry} and, more recently, as a function of materials and fabrication parameters.  These latter studies have aided in the reduction of the amplitude of $1/f$ flux noise in our devices and will be the subject of a forthcoming publication.  Using the methods described in Ref.~\onlinecite{1OverFGeometry}, we have generated the one-sided flux noise power spectral density $S_{\Phi}(f)$ shown in Fig.~\ref{fig:fluxnoise}.  These data have been fit to the generic form
\begin{equation}
\label{eqn:1OverF}
S(f)=\frac{A^2}{f^{\alpha}}+w_n\; ,
\end{equation}

\noindent with best fit parameters $\alpha=0.95\pm 0.05$, $\sqrt{w_n}=9.7\pm 0.5\,\mu\Phi_0/\sqrt{\text{Hz}}$ and amplitude $A$ such that $\sqrt{S_{\Phi}(1\,\text{Hz})}=1.3^{+0.7}_{-0.5}\,\mu\Phi_0/\sqrt{\text{Hz}}$.  Thus we reach the third key conclusion of this article:  We have demonstrated that it is possible to achieve $1/f$ flux noise levels with Nb wiring that are as low as the best Al wire qubits reported in the literature \cite{1OverF,1OverFFluxQubit1,1OverFFluxQubit2}.  Moreover, we have measured similar spectra from a large number of identical flux qubits, both on the same and different chips, and can state with confidence that the $1/f$ amplitude reported herein is reproducible.  Given the experimentally observed geometric scaling of $S_{\Phi}(1\,\text{Hz})$ in Ref.~\onlinecite{1OverFGeometry} and the relatively large size of our flux qubit bodies, we consider the prospects of observing even lower $1/f$ noise in smaller flux qubits from our fabrication facility to be very promising.

\subsection{Critical Current Noise}

A second noise metric of note is the critical current noise spectral density $S_I(f)$.  This quantity has been studied extensively and a detailed comparison of experimental results is presented in Ref.~\onlinecite{vanHarlingen}.  A recent study of the temperature and geometric dependence of critical current noise has been published in Ref.~\onlinecite{NewCriticalCurrentNoise}.  Based upon Eq.~(18) of Ref.~\onlinecite{vanHarlingen}, we estimate that the $1/f$ critical current noise from a single $0.6\,\mu$m diameter junction, as found in the CCJJ rf-SQUID flux qubit, will have an amplitude such that $\sqrt{S_I(1\,\text{Hz})}\sim 0.2\,$pA$/\sqrt{\text{Hz}}$.  Unfortunately, we were unable to directly measure critical current noise in the flux qubit.  While the QFP-enable readout provided high fidelity qubit state discrimination when qubits are fully annealed to $\Phiccjjx=-\Phi_0$, this readout mechanism simply lacked the sensitivity required for performing high resolution critical current noise measurements.  In lieu of a measurement of $S_I(f)$ from a qubit, we have characterized this quantity for the dc-SQUID connected to the qubit in question.  The dc-SQUID had two $0.6\,\mu$m junctions connected in parallel.  A time trace of the calibrated switching current $I_{\text{sw}}\approx I_c$ was obtained by repeating the waveform sequence depicted in Fig.~\ref{fig:unipolarannealing}b except with $\Philatchx=-\Phi_0/2$ at all time (QFP disabled, minimum persistent current) and $\Phi^x_{\text{ro}}=0$ to provide minimum sensitivity to flux noise.  Assuming that the critical current noise from each junction is uncorrelated, the best that we could establish was an upper bound of $\sqrt{S_I(1\,\text{Hz})}\lesssim 7\,$pA$/\sqrt{\text{Hz}}$ for a single $0.6\,\mu$m diameter junction.

Given the upper bound cited above for critical current noise from a single junction, we now turn to assessing the relative impact of this quantity upon the CCJJ rf-SQUID flux qubit.  It is shown in Appendix B that fluctuations in the critical currents of the individual junctions of a CCJJ generate apparent flux noise in the flux qubit by modulating $\Phi_q^0$.  Inserting critical current fluctuations of magnitude $\delta I_c\lesssim 7\,$pA$/\sqrt{\text{Hz}}$ and a mean junction critical current $I_c=I_q^c/4\sim 0.8\,\mu$A into Eq.~(\ref{eqn:4JOffsetFluctuation}) yields qubit degeneracy point fluctuations $\left|\delta\Phi_q^0\right|\lesssim 0.1\,\mu\Phi_0/\sqrt{\text{Hz}}$.  This final result is at least one order of magnitude smaller than the amplitude of $1/f$ flux noise inferred from the data in Fig.~\ref{fig:fluxnoise}.  As such, we consider the effects of critical current noise in the CCJJ rf-SQUID to be tolerable.

\subsection{Estimation of $T^*_{2}$}
 
While measurements of noise power spectral densities are the most direct way of reporting upon and comparing between different qubits, our research group is frequently asked what is the dephasing time for our flux qubits.  The answer presumably depends very strongly upon bias settings, for recall that we have measured properties of the CCJJ rf-SQUID flux qubit in both the coherent and incoherent regime.  Given that our apparatuses contain only low bandwidth bias lines for enabling AQO, we are unable to measure dephasing within our own laboratory.  Collaborative efforts to measure dephasing for our flux qubits are in progress.  In the meantime, we provide a rough estimate below for our flux qubits if they were biased to the optimal point, $\Phiqx=\Phi_q^0$ based upon the measured $S_{\Phi}(f)$ and subjected to a free induction decay, or Ramsey fringe, experiment.  Referring to Eq.~(33a) of Ref.~\onlinecite{Martinis} and key results from Ref.~\onlinecite{Schnirman}, the mean squared phase noise for a flux qubit at the optimal point  will be given by
\begin{equation}
\label{eqn:dephasing1}
\left<\phi_n^2(t)\right>=\frac{1}{\hbar^2}\frac{\left(2\iqp\right)^4}{2\Delta^2}\int^{\Delta/h}_{f_m}\! df S_{\Phi^2}(f)\frac{\sin^2(\pi f t)}{(\pi f)^2} \; ,
\end{equation}

\noindent where $S_{\Phi^2}(\omega)$ represents the quadratic flux noise spectral density and $f_m$ is the measurement cutoff frequency.  Assuming that the first order spectral density $S_{\Phi}(\omega)=2\pi A^2/\omega$, then $S_{\Phi^2}(\omega)$ can be written as 
\begin{eqnarray}
\label{eqn:sphisquared}
S_{\Phi^2}(\omega) & = & \frac{1}{2\pi}\int\! dt e^{-i\omega t}\left<\Phi_n^2(t) \Phi_n^2(0)\right> \nonumber\\
 & = &  \frac{1}{2\pi}\int\! dt e^{-i\omega t} \int d\omega^{\prime}\frac{2\pi A^2}{\omega^{\prime}}\int d\omega^{\prime\prime}\frac{2\pi A^2}{\omega^{\prime\prime}} \nonumber\\
  & = & 8\pi^2 A^4\frac{\ln\left(\omega/\omega_{\text{ir}}\right)}{\omega} \; ,
\end{eqnarray}

\noindent where $\omega_{\text{ir}}\equiv 2\pi f_{\text{ir}}$ denotes an infrared cutoff of the $1/f$ noise spectral density.  Inserting Eq.~(\ref{eqn:sphisquared}) into Eq.~(\ref{eqn:dephasing1}) and rendering the integral dimensionless then yields:
\begin{equation}
\label{eqn:dephasing2}
\left<\phi_n^2(t)\right>=\frac{t^2}{\hbar^2}\frac{\left(2\iqp A\right)^4}{\pi\Delta^2}\int^{\Delta t/h}_{f_{\text{min}}t}\! dx \frac{\ln\left(x/f_{\text{ir}}t\right)\sin^2(\pi x)}{x^3} \; ,
\end{equation}

\noindent where $f_{\text{min}}=\max\left[\begin{array}{cc} f_m & f_{\text{ir}}\end{array}\right]$.  We have numerically studied the behavior of the integral in Eq.~(\ref{eqn:dephasing2}).  In the very long measurement time limit the integral is cut off by $f_{\text{ir}}$ and the integral varies as $1/t^2$, which then cancels the factor of $t^2$ in the numerator of Eq.~(\ref{eqn:dephasing2}).  This means that the mean squared phase noise eventually reaches a finite limit.  However, the more experimentally relevant limit is $f_m\gg f_{\text{ir}}$ , for which we found empirically that the integral varies roughly as $5\times\left[\ln\left(f_m/f_{\text{ir}}\right)\right]^2$ over many orders of magnitude in the argument of the logarithm.  In this latter limit the result is independent of $t$, so Eq.~(\ref{eqn:dephasing2}) can be rewritten as $\left<\phi_n^2(t)\right>=t^2/(T^*_{2})^2$, which then yields the following formula for $T^*_{2}$:
\begin{equation}
\label{eqn:Tphi}
T^*_{2}\approx\left[\frac{1}{\hbar^2}\frac{\left(2\iqp A\right)^4}{\pi\Delta^2}5\ln\left(f_m/f_{\text{ir}}\right)\right]^{-1/2} \; .
\end{equation}

Since flux noise spectra seem to obey the $1/f$ form down to at least $0.1\,$mHz and researchers are generally concerned with dephasing over times of order $1\,\mu$s, then it is fair to consider $f_m/f_{\text{ir}}\sim 10^{10}$.  For a nominal value of $\Phiccjjx$ such that the flux qubit is in the coherent regime, say $-0.652\,\Phi_0$, the qubit parameters are $\Delta_q/h\approx 2\,$GHz and $\iqp\approx 0.7\,\mu$A.  Substituting these quantities into Eq.~(\ref{eqn:dephasing2}) then yields $T^*_{2}\sim 150\,$ns.  This estimate of the dephasing time is comparable to that observed in considerably smaller flux qubits with comparable $1/f$ flux noise levels \cite{1OverFFluxQubit1,1OverFFluxQubit2}.

\section{Conclusions}

One can draw three key conclusions from the work presented herein:  First, the CCJJ rf-SQUID is a robust and scalable device in that it allows for in-situ correction for parametric variations in Josephson junction critical currents and device inductance, both within and between flux qubits using only static flux biases.  Second, the measured flux qubit properties, namely the persistent current $\iqp$ and tunneling energy $\Delta_q$, agree with the predictions of a quantum mechanical Hamiltonian whose parameters have been independently calibrated, thus justifying the identification of this device as a flux qubit.  Third, it has been experimentally demonstrated that the low frequency flux noise in this all Nb wiring flux qubit is comparable to the best all Al wiring devices reported upon in the literature.  Taken in summation, these three conclusions represent a significant step forward in the development of useful large scale superconducting quantum information processors.

We thank J.~Hilton, P.~Spear, A.~Tcaciuc, F.~Cioata, M.~Amin, F.~Brito, D.~Averin, A.~Kleinsasser and G.~Kerber for useful discussions. Siyuan Han was supported in part by NSF Grant No. DMR-0325551.

\begin{appendix}

\section{CJJ rf-SQUID}

Let the qubit and cjj loop phases be defined as 
\begin{subequations}
\begin{equation}
\varphi_q\equiv\left(\varphi_1+\varphi_2\right)/2 \; ,
\end{equation} 
\begin{equation}
\varphi_{\text{cjj}}\equiv\varphi_1-\varphi_2 \; ,
\end{equation}
\end{subequations}
respectively.  Furthermore, assume that the CJJ loop has an inductance $L_{\text{cjj}}$ that is divided symmetrically between the two paths.  Using trigonometric relations, one can write a Hamiltonian for this system in terms of modes in the $q$ and cjj loops that has the following form:
\begin{subequations}
\begin{eqnarray}
\label{eqn:2J2DH}
{\cal H} & = & \sum_n\left[\frac{Q_n^2}{2C_n}+U_n\frac{(\varphi_n-\varphi_n^x)^2}{2}\right] \nonumber\\
 & & -U_q\beta_+\cos\left(\frac{\varphi_{\text{cjj}}}{2}\right)\cos\left(\varphi_q\right) \nonumber\\
  & & +U_q\beta_-\sin\left(\frac{\varphi_{\text{cjj}}}{2}\right)\sin\left(\varphi_q\right) \; ;
\end{eqnarray}
\vspace{-12pt}
\begin{equation}
\beta_{\pm}=\frac{2\pi L_q\left(I_{1}\pm I_{2}\right)}{\Phi_0} \; ,
\end{equation}
\end{subequations}

\noindent where the sum is over $n\in\left\{q,\text{cjj}\right\}$, $C_q\equiv C_1+C_2$, $1/C_{\text{cjj}}\equiv 1/C_1+1/C_2$, $U_n\equiv (\Phi_0/2\pi)^2/L_n$, $L_q\equiv L_{\text{body}}+L_{\text{cjj}}/4$ and $[\Phi_0\varphi_n/2\pi,Q_n]=i\hbar$.  The Josephson potential energy of Hamiltonian (\ref{eqn:2J2DH}) can be rearranged by defining an angle $\theta$ such that $\tan\theta=(\beta_-/\beta_+)\tan\left(\varphi_{\text{cjj}}/2\right)$.  Further trigonometric manipulation then yields Eqs.~(\ref{eqn:2JHeff})-(\ref{eqn:2Jbetapm}). 

\section{CCJJ rf-SQUID}

Following the same logic as for the CJJ rf-SQUID, one can define four orthogonal quantum mechanical degrees of freedom as follows:
\begin{subequations}
\begin{eqnarray}
\label{eqn:ccjjphases}
\varphi_L & \equiv & \varphi_1-\varphi_2 \; ;\\
\varphi_R & \equiv & \varphi_3-\varphi_4 \; ;\\
\varphi_{\text{ccjj}} & \equiv & \varphi_{\ell}-\varphi_r=\frac{\varphi_1+\varphi_2}{2}-\frac{\varphi_3+\varphi_4}{2} \; ;\\
\varphi_q & \equiv & \frac{\varphi_{\ell}+\varphi_r}{2}=\frac{\varphi_1+\varphi_2+\varphi_3+\varphi_4}{4}\; .
\end{eqnarray}
\end{subequations}

\noindent  Using the same strategy as in Appendix A, one can use trigonometric identities to first express the Josephson potential in terms of the $L$ and $R$ loop modes:
\begin{subequations}
\begin{eqnarray}
\label{eqn:4J4DH}
{\cal H} & = & \sum_n\frac{Q_n^2}{2C_n}+\sum_mU_m\frac{(\varphi_m-\varphi_m^x)^2}{2} \nonumber\\
 & & -U_q\beta_{L+}\cos\left(\frac{\varphi_L}{2}\right)\cos\left(\varphi_{\ell}\right) \nonumber\\
  & & +U_q\beta_{L-}\sin\left(\frac{\varphi_L}{2}\right)\sin\left(\varphi_{\ell}\right) \nonumber \\
 & & -U_q\beta_{R+}\cos\left(\frac{\varphi_R}{2}\right)\cos\left(\varphi_{r}\right) \nonumber\\
  & & +U_q\beta_{R-}\sin\left(\frac{\varphi_R}{2}\right)\sin\left(\varphi_{r}\right) \; ;
\end{eqnarray}
\vspace{-12pt}
\begin{equation}
\label{eqn:betalrpm}
\beta_{L(R),\pm}\equiv\frac{2\pi L_q\left(I_{1(3)}\pm I_{2(4)}\right)}{\Phi_0} \; ,
\end{equation}
\end{subequations}

\noindent where the first sum is over $n\in\left\{L,R,\ell,r\right\}$ and the second sum is over closed inductive loops $m\in\left\{L,R,\text{ccjj},q\right\}$.  As before, each of the modes obey the commutation relation $[\Phi_0\varphi_n/2\pi,Q_n]=i\hbar$.  Here, $1/C_{L(R)}=1/C_{1(3)}+1/C_{2(4)}$, $C_{\ell(r)}=C_{1(3)}+C_{2(4)}$ and $U_m=(\Phi_0/2\pi)^2/L_m$. 

We have found it adequate for our work to assume that $L_{L,R}/L_q\ll 1$, which then allows one to reduce the four dimensional system given in Hamiltonian (\ref{eqn:4J4DH}) to two dimensions.  Consequently, we will substitute $\varphi_{L(R)}=\varphi^x_{L(R)}$ and ignore the $L$ and $R$ kinetic terms henceforth.  Assuming that the inductance of the ccjj loop is divided equally between the two branches one can then write $L_q=L_{\text{body}}+L_{\text{ccjj}}/4$.  With these approximations and the $\theta$ strategy presented in Appendix A, one can rearrange the Josephson potential terms to yield the following:
\begin{subequations}
\begin{eqnarray}
\label{eqn:4J2DHver1}
{\cal H} & = & \sum_n\frac{Q_n^2}{2C_n}+\sum_mU_m\frac{(\varphi_m-\varphi_m^x)^2}{2} \nonumber\\
 & & -U_q\beta_{L}\cos\left(\varphi_{\ell}-\varphi_L^0\right) \nonumber\\
  & & -U_q\beta_{R}\cos\left(\varphi_{r}-\varphi_R^0\right) \; ;
\end{eqnarray}
\vspace{-12pt}
\begin{eqnarray}
\label{eqn:betalr}
\beta_{L(R)}  & = & \beta_{L(R),+}\cos\left(\frac{\varphi_{L(R)}^x}{2}\right) \\
 & & \times\sqrt{1+\left[\frac{\beta_{L(R),-}}{\beta_{L(R),+}}\tan\left(\frac{\varphi_{L(R)}^x}{2}\right)\right]^2} \; ; \nonumber
\end{eqnarray}
\vspace{-12pt}
\begin{equation}
\label{eqn:4JMinorOffset}
\varphi_{L(R)}^0 
 =-\arctan\left(\frac{\beta_{L(R),-}}{\beta_{L(R),+}}\tan(\varphi_{L(R)}^x/2)\right)\; ,
\end{equation}
\end{subequations}

\noindent where the first sum is over $n\in\left\{\ell,r\right\}$ and the second sum is over $m\in\left\{\text{ccjj},q\right\}$.  The Josephson potential is given by a sum of two cosines, as encountered in the CJJ rf-SQUID derivation of Hamiltonian (\ref{eqn:2J2DH}) from Hamiltonian (\ref{eqn:Hphase}).  These two terms can be rewritten in the same manner by defining $\beta_{\pm}=\beta_L\pm\beta_R$.  The result, similar to Hamiltonian (\ref{eqn:2J2DH}), can then be subjected to the $\theta$ strategy to yield
\begin{subequations}
\begin{eqnarray}
\label{eqn:4J2DHver2}
{\cal H} &  = & \sum_n\left[\frac{Q_n^2}{2C_n}+U_n\frac{(\varphi_n-\varphi_n^x)^2}{2}\right] \nonumber\\
 & & -U_q\beta_{\text{eff}}\cos\left(\varphi_q-\varphi_q^0\right) \; ,
 \end{eqnarray}
 
\noindent where the sum is over $n\in\left\{q,\text{ccjj}\right\}$ and the capacitances are defined as $C_q=C_1+C_2+C_3+C_4$ and $1/C_{\text{ccjj}}=1/(C_1+C_2)+1/(C_3+C_4)$.  The other parameters are defined as
\begin{equation}
\label{eqn:4JBeff}
\beta_{\text{eff}}=\beta_+\cos\left(\frac{\gamma}{2}\right)\sqrt{1+\left[\frac{\beta_-}{\beta_+}\tan\left(\frac{\gamma}{2}\right)\right]^2} \; ;
\end{equation}
\vspace{-12pt}
\begin{equation}
\label{eqn:4JQOffset}
\varphi_q^0=\frac{\varphi_L^0+\varphi_R^0}{2}+\gamma_0 \; ;
\end{equation}
\vspace{-12pt}
\begin{equation}
\label{eqn:4JCCJJPhase}
\gamma\equiv\varphi_{\text{ccjj}}-\left(\varphi_L^0-\varphi_R^0\right) \; ;
\end{equation}
\vspace{-12pt}
\begin{equation}
\label{eqn:4JCCJJMonkey}
\gamma_0\equiv -\arctan\left(\frac{\beta_-}{\beta_+}\tan(\gamma/2)\right) \; ;
\end{equation}
\begin{equation}
\label{eqn:betaccjjpm}
\beta_{\pm}\equiv \beta_L\pm\beta_R \; .
\end{equation}
\end{subequations}

Hamiltonian (\ref{eqn:4J2DHver2}) inherits much of its complexity from junction asymmetry both within the minor loops, which gives rise to $\varphi_{L(R)}^0$, and effective junction asymmetry between the minor loops, which gives rise to $\gamma_0$.  For arbitrary external flux biases and nominal spread in junction critical current, the CCJJ rf-SQUID offers no obvious advantage over the CJJ rf-SQUID.  However, upon choosing biases $\Phi_L^x$ and $\Phi_R^x$ such that 
\begin{equation}
\label{eqn:balanced}
\beta_L=\beta_R \;\; ,
\end{equation}

\noindent then $\beta_-=0$, and consequently $\gamma_0=0$.  With these substitutions, Hamiltonian (\ref{eqn:4J2DHver2}) yields Hamiltonian (\ref{eqn:4JHeff}).  Note that for $\beta_{L(R),-}/\beta_{L(R),+}\ll 1$ and modest $\Phi^x_{L(R)}$ that the so-called CCJJ balancing condition given by Eqs.~(\ref{eqn:betalr}) and (\ref{eqn:balanced}) can be written approximately as
\begin{displaymath}
\beta_{L,+}\cos\left(\frac{\varphi_L^x}{2}\right) \approx \beta_{R,+}\cos\left(\frac{\varphi_R^x}{2}\right) \; ,
\end{displaymath}

\noindent which, upon solving for $\varphi_L^x$ yields
\begin{equation}
\label{eqn:balancedapprox}
\Philx  =  \frac{2\pi}{\Phi_0}\arccos\left[\frac{\beta_{R,+}}{\beta_{L,+}}\cos\left(\frac{\pi\Phirx}{\Phi_0}\right)\right] \; .
\end{equation}

It is possible for critical current noise to couple into the $\varphi_q$ degree of freedom in any compound junction rf-SQUID qubit via modulation of the junction asymmetry-dependent apparent qubit flux offset $\Phi_q^0$.  In the case of the CCJJ rf-SQUID, all three quantities on the right side of Eq.~(\ref{eqn:4JQOffset}) are ultimately related to the critical currents of the individual junctions.  Given typical junction parameter spreads from our fabrication facility,
\begin{displaymath}
\left|\frac{\beta_{L(R),-}}{\beta_{L(R),+}}\right|=\left|\frac{I_{1(3)}-I_{2(4)}}{I_{1(3)}+I_{2(4)}}\right|\sim {\cal O}(0.01) \; ,
\end{displaymath}

\noindent so one can write an approximate expression for $\varphi^0_{L(R)}$ using Eq.~(\ref{eqn:4JMinorOffset}):
\begin{eqnarray}
\label{eqn:4JMinorOffsetApprox}
\varphi^0_{L(R)} & \approx & -\frac{I_{1(3)}-I_{2(4)}}{I_{1(3)}+I_{2(4)}}\tan\left(\frac{\varphi^x_{L(R)}}{2}\right) \nonumber\\
 & \approx &  -\frac{I_{1(3)}-I_{2(4)}}{2I_c}\tan\left(\frac{\varphi^x_{L(R)}}{2}\right) \; ,
\end{eqnarray}

\noindent and for $\gamma_0$ using Eqs.~(\ref{eqn:betalr}), (\ref{eqn:4JCCJJPhase}) and (\ref{eqn:4JCCJJMonkey}):
\begin{eqnarray}
\label{eqn:4JCCJJMonkeyApprox}
\gamma_0 & \approx & \frac{(I_3+I_4)\cos\left(\frac{\varphi_R^x}{2}\right)-(I_1+I_2)\cos\left(\frac{\varphi_L^x}{2}\right)}{(I_1+I_2)\cos\left(\frac{\varphi_L^x}{2}\right)+(I_3+I_4)\cos\left(\frac{\varphi_R^x}{2}\right)}\tan\left(\frac{\gamma}{2}\right) \nonumber\\
 & \approx & \frac{(I_3+I_4)\cos\left(\frac{\varphi_R^x}{2}\right)-(I_1+I_2)\cos\left(\frac{\varphi_L^x}{2}\right)}{2I_c\left[\cos\left(\frac{\varphi_L^x}{2}\right)+\cos\left(\frac{\varphi_R^x}{2}\right)\right]}\tan\left(\frac{\gamma}{2}\right) , \nonumber\\
  & & 
\end{eqnarray}

\noindent where $I_c$ represents the mean critical current of a single junction.  The CCJJ rf-SQUID is intended to be operated with only small flux biases in the minor loops, thus $\cos\left(\frac{\varphi_L^x}{2}\right)\approx\cos\left(\frac{\varphi_R^x}{2}\right)\approx 1$.  It is also reasonable to assume that $\gamma\approx\varphi^x_{\text{ccjj}}$
as the corrections to $\tan(\gamma/2)$ from $\varphi^0_{L(R)}$ and from the effective two-dimensionality of the rf-SQUID potential will be very small.  Inserting Eqs.~\ref{eqn:4JMinorOffsetApprox} and \ref{eqn:4JCCJJMonkeyApprox} into Eq.~(\ref{eqn:4JQOffset}) then yields
\begin{eqnarray}
\label{eqn:4JOffsetApprox}
\varphi^0_q & \approx & -\frac{I_1}{2I_c}\left[\tan\left(\frac{\varphi_L^x}{2}\right)+\frac{1}{2}\tan\left(\frac{\varphi^x_{\text{ccjj}}}{2}\right)\right] \nonumber\\
 & &  -\frac{I_2}{2I_c}\left[-\tan\left(\frac{\varphi_L^x}{2}\right)+\frac{1}{2}\tan\left(\frac{\varphi^x_{\text{ccjj}}}{2}\right)\right] \nonumber\\
 & &  -\frac{I_3}{2I_c}\left[\tan\left(\frac{\varphi_R^x}{2}\right)-\frac{1}{2}\tan\left(\frac{\varphi^x_{\text{ccjj}}}{2}\right)\right] \nonumber\\
& &  -\frac{I_4}{2I_c}\left[-\tan\left(\frac{\varphi_R^x}{2}\right)-\frac{1}{2}\tan\left(\frac{\varphi^x_{\text{ccjj}}}{2}\right)\right] \; .
\end{eqnarray}

For the typical operating parameters described in this article, $\Phi^x_{L(R)}/\Phi_0\sim0.1$ and the device acts as a qubit for $\Phiccjjx/\Phi_0\sim 0.65$.  For these flux biases, the magnitude of the terms within the square braces in Eq.~(\ref{eqn:4JOffsetApprox}) are all of order 1.  Therefore, for general flux bias conditions, the apparent qubit flux offset is roughly given by
\begin{displaymath}
\Phi_q^0 \approx -\frac{\Phi_0}{4\pi}\frac{(I_1+I_2)-(I_3+I_4)}{I_c} \; .
\end{displaymath}

\noindent Assume that each junction experiences critical current fluctuations of magnitude $\delta I_c$.  If each junction's fluctuations are independent, 
then the root mean square variation of the qubit degeneracy point $\left|\delta\Phi_q^0\right|$ will be 
\begin{equation}
\label{eqn:4JOffsetFluctuation}
\left|\delta\Phi_q^0\right| \approx \frac{\Phi_0}{2\pi}\frac{\delta I_c}{I_c} \; .
\end{equation}

\noindent Thus, critical current fluctuations generate apparent flux noise in the CCJJ rf-SQUID flux qubit.

\end{appendix}

\bibliography{CCJJQubit}

\end{document}